\documentclass[%
 aps,
pra,
 amsmath,amssymb,
superscriptaddress,
altaffilletter
]{revtex4-2}
\usepackage[T1]{fontenc}
\usepackage[latin9]{inputenc}
\usepackage{geometry}
\usepackage{amssymb}
\usepackage{graphicx}
\usepackage{esint}
\usepackage{hyperref}
\usepackage{lineno}
\newcommand{\etal}{{\it et al.}}
\newcommand{\Arlington}{\affiliation{Department of Physics, University of Texas at Arlington, Arlington, TX 76019, USA}}

\newcommand{\Mainz}{\affiliation{Institute for Physics, Johannes Gutenberg University Mainz, 55128 Mainz, Germany}}
\newcommand{\MIT}{\affiliation{Laboratory for Nuclear Science, Massachusetts Institute of Technology, Cambridge, MA 02139, USA}}
\newcommand{\PennStateARL}{\affiliation{Applied Research Laboratory, Pennsylvania State University, University Park, PA 16802, USA}}
\newcommand{\PennState}{\affiliation{Department of Physics, Pennsylvania State University, University Park, PA 16802, USA}}
\newcommand{\PNNL}{\affiliation{Pacific Northwest National Laboratory, Richland, WA 99354, USA}}
\newcommand{\Yale}{\affiliation{Wright Laboratory and Department of Physics, Yale University, New Haven, CT 06520, USA}}
\newcommand{\Livermore}{\affiliation{Lawrence Livermore National Laboratory, Livermore, CA 94550, USA}}
\newcommand{\Case}{\affiliation{Department of Physics, Case Western Reserve University, Cleveland, OH 44106, USA}}
\newcommand{\Heidelberg}{\affiliation{Institute for Theoretical Astrophysics, Heidelberg University, 69120 Heidelberg, Germany}}
\newcommand{\Illinois}{\affiliation{Department of Physics, University of Illinois Urbana-Champaign, Urbana, IL 61801, USA}}
\newcommand{\Indiana}{\affiliation{Center for Exploration of Energy and Matter and Department of Physics, Indiana University, Bloomington, IN, 47405, USA}}
\newcommand{\KIT}{\affiliation{Institute of Astroparticle Physics, Karlsruhe Institute of Technology, 76021 Karlsruhe, Germany}}
\newcommand{\Pitt}{\affiliation{Department of Physics \& Astronomy, University of Pittsburgh, Pittsburgh, PA 15260, USA}}
\newcommand{\Washington}{\affiliation{Center for Experimental Nuclear Physics and Astrophysics and Department of Physics, University of Washington, Seattle, WA 98195, USA}}
\newcommand{\Ghent}{\affiliation{Department of Physics and Astronomy, Ghent University, 9000 Ghent, Belgium}}

\makeatletter

\providecommand{\tabularnewline}{\\}

\@ifundefined{date}{}{\date{}}
\makeatother
\usepackage{babel}
\begin{document}
\title{Dynamics of Magnetic Evaporative Beamline Cooling for Preparation of Cold Atomic Beams}
\author{A.~Ashtari~Esfahani}\Washington
\author{S.~Bhagvati}\PennState
\author{S.~B\"oser}\Mainz
\author{M.~J.~Brandsema}\PennStateARL
\author{R.~Cabral}\Indiana
\author{V.~A.~Chirayath}\Arlington
\author{C.~Claessens}\Washington
\author{N.~Coward}\Arlington
\author{L.~de~Viveiros}\PennState
\author{P.~J.~Doe}\Washington
\author{M.~G.~Elliott}\Arlington
\author{S.~Enomoto}\Washington
\author{M.~Fertl}\Mainz
\author{J.~A.~Formaggio}\MIT
\author{B.~T.~Foust}\PNNL
\author{J.~K.~Gaison}\PNNL
\author{P.~Harmston}\Illinois
\author{K.~M.~Heeger}\Yale
\author{B.~J.~P.~Jones}
\email[Corresponding author: ]{ben.jones@uta.edu}
\Arlington
\author{E.~Karim}\Pitt
\author{K.~Kazkaz}\Livermore
\author{P.~T.~Kolbeck}\Washington
\author{M.~Li}\MIT
\author{A.~Lindman}\Mainz
\author{C.-Y.~Liu}\Illinois
\author{C.~Matth\'e}\Mainz
\author{R.~Mohiuddin}\Case
\author{B.~Monreal}\Case
\author{B.~Mucogllava}\Mainz
\author{R.~Mueller}\PennState
\author{A.~Negi}\Arlington
\author{J.~A.~Nikkel}\Yale
\author{E.~Novitski}\Washington
\author{N.~S.~Oblath}\PNNL
\author{M.~Oueslati}\Indiana
\author{J.~I.~Pe\~na}\MIT
\author{W.~Pettus}\Indiana
\author{V.~S.~Ranatunga}\Case
\author{R.~Reimann}\Mainz
\author{A.~L.~Reine}\Indiana
\author{R.~G.~H.~Robertson}\Washington
\author{L.~Salda\~na}\Yale
\author{P.~L.~Slocum}\Yale
\author{F.~Spanier}\Heidelberg
\author{J.~Stachurska}\Ghent
\author{K.~Stogsdill}\Arlington
\author{Y.-H.~Sun}\Washington
\author{P.~T.~Surukuchi}\Pitt
\author{L.~Taylor}\Arlington
\author{A.~B.~Telles}\Yale
\author{F.~Thomas}\Mainz
\author{L.~A.~Thorne}\Mainz
\author{T.~Th\"ummler}\KIT
\author{W.~Van~De~Pontseele}\MIT
\author{B.~A.~VanDevender}\Washington\PNNL
\author{T.~E.~Weiss}\Yale
\author{M.~Wynne}\Washington
\author{A.~Ziegler}\PennState
\collaboration{Project 8 Collaboration}\noaffiliation 
 

\date{01/17/2025}
\begin{abstract}
The most sensitive direct neutrino mass searches today are based on
measurement of the endpoint of the beta spectrum of tritium to infer
limits on the mass of the unobserved  neutrino. To avoid
the smearing associated with the distribution of molecular final states
in the T-He molecule, the next generation of these experiments will
need to employ atomic (T) rather than molecular (T$_{2}$) tritium
sources, at currents of at least 10$^{15}$ atoms per second. Following production, atomic T can be trapped in gravitational
and/or magnetic bottles for beta spectrum experiments, if and only if
it can first be cooled to millikelvin temperatures. Accomplishing this cooling
presents substantial technological challenges. The Project 8 collaboration
is developing a technique based on magnetic evaporative cooling along
a beamline (MECB) for the purpose of cooling T to feed a magnetogravitational
trap that also serves as a cyclotron radiation emission spectroscope.  Initial tests of the approach are planned in a pathfinder apparatus using atomic Li.
This paper presents a method for analyzing the dynamics of the MECB
technique, and applies these calculations to the design of systems for cooling and slowing of atomic Li and T. A
scheme is outlined that could provide a current of T at the millikelvin temperatures required for the Project 8 neutrino mass search.

\end{abstract}

\maketitle
\section{Introduction }

The unknown absolute value of the mass of the neutrino is one of the
most glaring holes in our understanding of particle physics today.
Neutrino oscillations
have proven definitively that at least two of the neutrino mass eigenstates
have non-zero masses~\cite{Super-Kamiokande:1998kpq,SNO:2002tuh,gonzalez2021nufit}, setting lower limits
on the electron-flavor-weighted neutrino mass $m_{\beta}$ in beta decay
searches at 9 meV (normal mass ordering) and 48 meV (inverted mass
ordering) respectively~\cite{Workman:2022ynf}.  The actual masses may be larger than these limits and the ordering is also unknown.  The most sensitive direct search for neutrino mass to date is the
currently running KATRIN experiment, which reconstructs the
energies of electrons emitted in beta decays of molecular tritium (T$_{2}$) using a Magnetic Adiabatic Collimation combined with Electrostatic (MAC-E)
filter \cite{Aker:2021ty}. This technique establishes the endpoint
shape of the beta spectrum, yielding the current world-leading direct
upper limit of $m_{\beta}\leq0.45\,\mathrm{eV}$ at 90\% confidence level (CL) \cite{aker2024direct}.  Neutrinoless double beta decay
is also a probe of neutrino mass, if and only if the neutrino is a
Majorana particle~\cite{dolinski2019neutrinoless}, although given uncertainties about the neutrino nature, mass ordering, and mixing parameters, without
an observation no upper limit can be drawn. Cosmology
is sensitive to the sum of the neutrino masses. Within the framework of the standard dark energy with cold dark matter ($\Lambda$CDM) model upper limits below the
KATRIN bound \cite{Aghanim:2020aa} and even below the limits set by oscillations~\cite{DESI:2025ejh} are found.  The current situation heightens the need for still more sensitive direct measurements below  the present limits from KATRIN. Since
KATRIN has effectively saturated the power of both the MAC-E filtering
technique and of T$_{2}$, paradigm-shifting new technologies are
required to enable further progress.

One promising method to advance the precision of neutrino mass measurements
in the laboratory is cyclotron radiation emission spectroscopy (CRES)
of electrons produced in decays of tritium~\cite{Monreal:2009za,Esfahani:2017dmu,Project8:2022hun,PhaseIIPRC}. In this
technique, electron energy is accessed with great precision by measuring
cyclotron frequencies in a magnetic trapping field. To achieve sensitivities
significantly below the KATRIN goal of $m_{\beta}\leq0.3\,\mathrm{eV}$ at 90\% CL, 
CRES must be complemented by an atomic T source, since T$_{2}$ introduces
irreducible uncertainties associated with excitation of
molecular final states \cite{Bodine:2015aa,Schneidewind:2023xmj}. The Project 8 target is 40~meV at 90\% CL.  Realization of a high
throughput atomic source, delivering a sufficient atom current to enable a sensitive neutrino mass measurement through beta decay endpoint studies of magnetically trapped T, is one of the primary R\&D goals of Project
8 in the next decade.

The proposed Project 8 trap is vertically oriented and closed magnetically at its lower extreme
to hold atoms gravitationally in the upward direction. Magnetogravitational traps have been used successfully for containment of ultra-cold neutrons~\cite{walstrom2009magneto,pattie2018measurement}. 
The T cooling stage must operate
continuously in the atomic beamline, fed by a source of initially
hot T atoms. Multiple source configurations are under investigation.
Approaches based on thermal cracking produce an initial flux at 2200
K~\cite{tschersich1998formation}, whereas electron cyclotron resonance
sources~\cite{Alton_10.1063/1.1144954} may provide a somewhat more
modest temperature. In either case, an intermediate
stage of cooling can then be performed by scattering T off cold
surfaces in a process known as thermal accommodation to between 10 and 30
K, below which temperature the recombined T$_2$ vapor will freeze to solid surfaces and be taken out of circulation. In
previous experiments with hydrogen, further cooling has been achieved
by down-scattering the atomic vapor from a film of superfluid liquid
helium \cite{silvera1982stabilization}; however, in the case of tritium
this technique is not expected to be effective, as the larger physisorption
energy indicates that recombination will be unavoidable at the T-He
interface \cite{WOS:A1979HA22200010}.  Though we lack direct experimental confirmation that this approach will not work for T, the physisorption concern prompts us to explore other methods of cooling and trapping as the Project 8 baseline.  A crucial technical step is
the further cooling of a large flux of moderately cold atomic T to
millikelvin temperatures with high throughput, minimal recombination, high purity, and continuous operation.

The Project 8 collaboration is pursuing R\&D on a Magnetic Evaporative
Cooling Beamline (MECB) for this purpose. This scheme is based on
the pioneering work of Hess~\cite{hess1986evaporative,masuhara1988evaporative}
that eventually led to successful Bose Einstein condensation of atomic
H~\cite{fried1998bose}.  For Project 8, evaporative cooling is implemented
continuously as a function of the axial coordinate $z$ of a linear atom guide, rather than in a single position
as a function of time, as employed in cold atom
traps~\cite{ketterle1996evaporative}.  Under the MECB scheme, low-field-seeking
spin polarized atoms are confined radially within a cylindrical magnetic
potential well. The highest energy atoms are selectively rejected
as the cloud travels along the beamline. Assuming the population
remains in thermal equilibrium, the escape of relatively energetic atoms leads to cooling.
Unlike in the case of atom traps, this method here does not require pre-cooling with either superfluid He surfaces or with laser cooling. The latter is ineffective for T, because available continuous-wave
lasers do not exist with sufficient power at the challenging 121-nm
wavelength associated with the Lyman alpha line, the only optically accessible
spectral feature from the T ground state.  Devices exhibiting MECB-like dynamics but
using radio frequency excitation to perform the evaporation cut
have been discussed for heavier atoms, see, for example, Refs.~\cite{lahaye2005evaporative,mandonnet2000evaporative}.
Application to atomic T carries unique additional challenges that
will be elaborated below.

One of the demonstrator phases en route to this system is a cold atomic
lithium (Li) beamline, providing the opportunity to prove the technique
 with a more convenient atomic
source. This technology development platform is advantageous, since Li is non-radioactive, experiences similar magnetic trapping forces to T, and is laser addressable via the D lines for density mapping and Doppler thermometry to monitor thermalization,  cooling and slowing.   A tunable flux of pre-cooled
$^{6}$Li will be delivered from an oven via a Zeeman
slower \cite{bowden2016adaptable,garwood2022hybrid,marti2010two}, enabling beam parameters to be tuned to replicate those
expected to emerge from the T accommodator in the final Project 8 beamline.
A key goal of this pathfinder experiment is to validate quantitative
models of MECB dynamics. The tools we describe in this paper have
been developed both to study the performance of a possible MECB implementation
using atomic T for Project 8, and to provide design specifications
for the atomic $^{6}$Li pathfinder project and predictions for its performance. 

The approaches presented in this work are not the only conceivable schemes for slowing a fast-moving atomic beam of T.   An ingenious method for slowing polar molecules in a spinning electrostatic guide has been developed by Chervenkov \etal~\cite{chervenkov2014} and a similar approach is possible for magnetic atoms.  H and D have been slowed using multi-stage Zeeman decelerators~\cite{hogan2007zeeman} and atomic coil gun geometries~\cite{libson2012atomic}.  However, for any of these approaches, operating them at the very high atom currents needed for the Project 8 application and in the difficult environment of a tritium handling facility remains to be demonstrated and appears to be very technically challenging.

In what follows, we outline a new method of analysis of MECB dynamics
(Sec. \ref{sec:Theory-of-Magnetic}). We discuss entrainment of magnetic atoms in the relevant
magnetic multipole fields (Sec \ref{subsec:Trapping-in-magnetic}),
the dynamics of evaporative beam cooling (Sec \ref{subsec:Cooling-dynamics})
including the details of cross sections  (Sec \ref{subsec:Triplet-scattering-cross})
and collision calculations (Sec \ref{subsec:Beam-cooling-via}).  This is followed
by an analysis of the proposed slowing method via transverse perturbation
(Sec \ref{subsec:Beam-slowing-via}). These tools are then used
to predict the performance of a MECB geometry designed for $^{6}$Li
cooling and slowing (Sec \ref{sec:MECB-Geometry-Proposal}) and finally
a speculative atomic T cooling
MECB geometry for Project 8 (Sec \ref{sec:A-speculative-geometry}).
Our conclusions are in Sec \ref{sec:Conclusions}.

\section{The magnetic evaporative cooling beamline concept\label{sec:Theory-of-Magnetic}}

In this section we review the concepts central to the MECB for Project 8, in particular, methods of laser-less
trapping, cooling and slowing in linear magnetic multipole geometries.  For reference, a table of variable names can be found in \hyperref[sec:AppendixB]{Appendix B}.

\subsection{Trapping in magnetic multipole guides \label{subsec:Trapping-in-magnetic}}
\begin{figure}
\begin{centering}
\includegraphics[width=1\columnwidth]{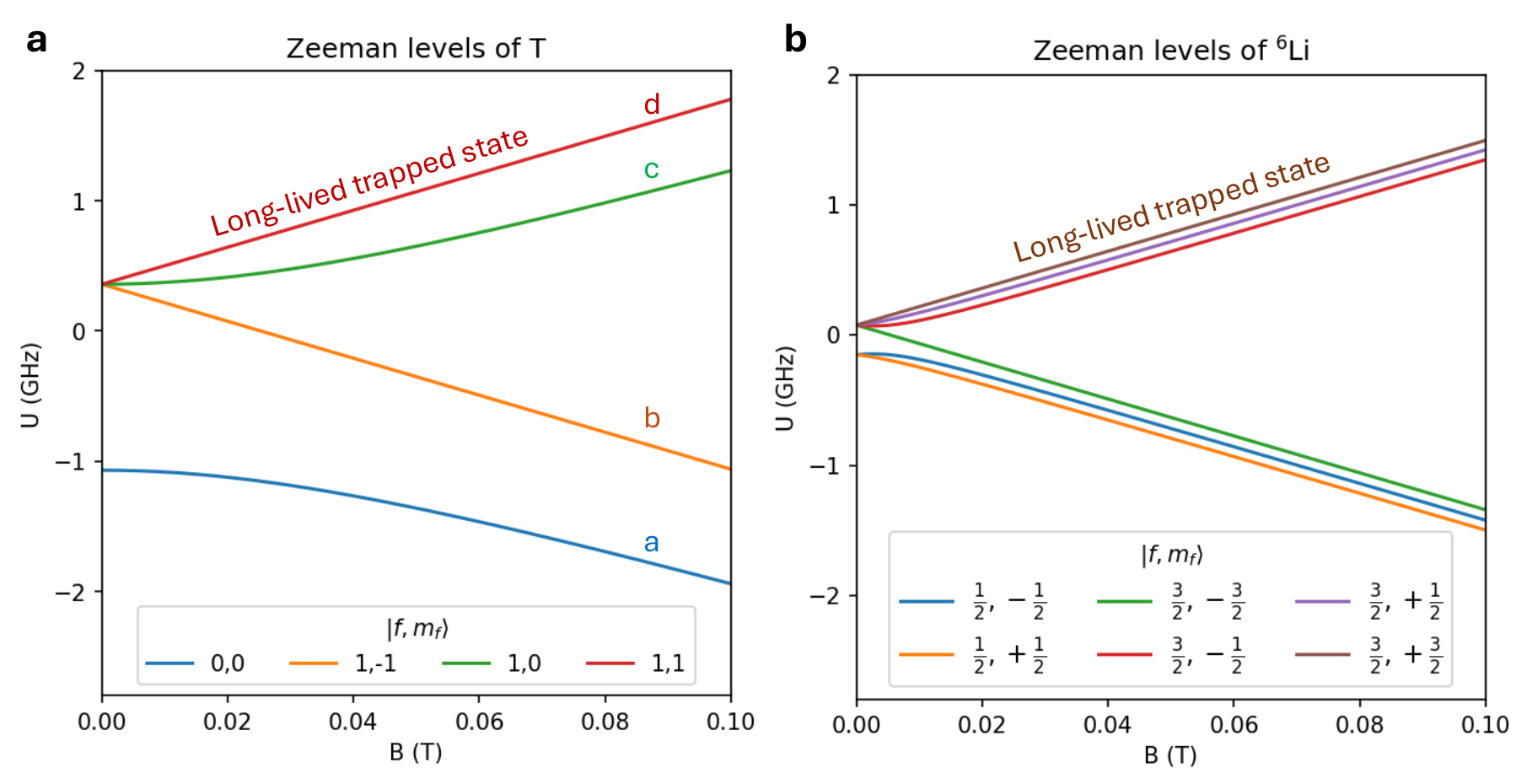}
\par\end{centering}
\caption{Hyperfine levels of (a) atomic T;  and (b) atomic $^{6}$Li.  Only the fully stretched states, labeled as ``Long-lived trapped state'' is expected to survive for a prolonged time in the magnetic beamline geometry, as the high-field-seekers are quickly ejected from the beam and the remaining low-field-seekers will be lost through spin-exchange collisions. \label{fig:Hyperfine-levels-of}}
\end{figure}

Investigations into spin polarized atomic H and $^{6}$Li have been
widespread in AMO physics~\cite{schunemann1998magneto,mosk2001resonator,mosk2001mixture,tiecke2009high,vinogradov2020trapping}. Trapping of these species relies
on the interaction of the atom, which in each case has a single unpaired
outer electron, with applied magnetic fields. In both of these atoms,
the nucleus also has a non-zero spin.
In the case of H or T, the nucleus has spin 1/2, and the nuclear and electron spins can therefore couple
in either a singlet ($f=0$) or a triplet ($f=1$) configuration. This coupling
leads to two possible hyperfine levels in a zero magnetic field. As depicted in Fig.~\ref{fig:Hyperfine-levels-of}, in a non-zero field, these levels are further split by the Zeeman effect, since
the interaction between the electron and nucleus with the external
 field generates inequivalent energy perturbations, in each case proportional
to the magnetic moments $\mu_{i}=e/2m_{i}$ where $m_{i}$ is the mass of the electron
(e) or nucleus (N), respectively. The four resulting Zeeman-split
levels of hydrogen and tritium are commonly denoted in the literature as $a,b,c,d$, from
least to most energetic. In the case of $^{6}$Li, the nucleus has
spin 1. There are still two hyperfine levels at zero field, this
time corresponding to doublet ($f=1/2$) and quartet ($f=3/2$) total
spin configurations, and applied $B$ fields split this into a manifold
of 6 Zeeman levels. The Zeeman-split energy levels can
be calculated nonperturbatively using the Breit-Rabi method, diagonalizing
the relevant Hamiltonian to account consistently for both Zeeman and
hyperfine effects, as shown in Fig.~\ref{fig:Hyperfine-levels-of}.
The eigenstates can be labeled as $|f,m_{f}\rangle$, the states to which they are adiabatically connected as $B\rightarrow0$.

In a large magnetic field the Zeeman perturbation dominates over the
hyperfine one, and because $\mu_{e}\gg\mu_{N}$ it is the coupling
of the electron that is dominant in determining the energy of
a given level. The states thus split into two groups: the so-called
``high-field seekers'' where the electron spin is predominantly anti-aligned (and hence magnetic moment aligned)
with the magnetic field, whose energies fall as the field strength
increases; and ``low-field seekers'' where the electron spin is predominantly
aligned with the field, whose energies increase with field strength. Magnetic fields can admit static minima in free space but not
static maxima, and as such only the low-field-seeking states can be
trapped in a time-invariant magnetic potential. Because the 
unpaired electron determines the trap strength, the depths of a given
potential well for low-field seekers of T, $^{6}$Li, or indeed
any other alkali metal are remarkably similar. 

Although all low-field-seeking $|f,m_{f}\rangle$ states can be trapped,
not all can be expected to exhibit a long trap lifetime. In the case
of T, the $c$ state (which adiabatically connects to $|1,0\rangle$)
is composed of electron and nuclear spin states as $|c\rangle=\cos \theta |\uparrow_{e}\rangle|\downarrow_{N}\rangle+\sin \theta|\downarrow_{e}\rangle|\uparrow_{N}\rangle$,
with a mixing parameter $\theta$ that depends on $B$ field~\cite{silvera1981spin}. These states can be lost through spin-exchange collisions $cc\rightarrow ca$, $cc\rightarrow ac$, and $cc\rightarrow bd$~\cite{lagendijk1986spin}. After a significant time period therefore, only $d$
states are expected to remain in a magnetic trap. A similar argument
suggests that in the case of $^{6}$Li, the state analytically continuing
to $|\frac{3}{2},\frac{3}{2}\rangle$ is the one that can be expected
to enjoy a long trap lifetime. These trapped states are labeled in
Fig. \ref{fig:Hyperfine-levels-of}.

The discussion so far has suggested magnetic fields varying in magnitude
but always aligned or anti-aligned with the atomic spin. A moving
atom, however, can traverse magnetic fields pointing
in different directions. In geometries where the $B$ field direction is
changing relatively slowly in space, the adiabatic theorem determines that
the spin state rotates to follow the direction of the field~\cite{kato1950adiabatic}.
As such, provided that the adiabatic condition is met, it is only the magnitude
of a magnetic field, not its direction, that determines the trapping
potential for low-field-seekers. As a consequence
of this principle, effective atom traps can be made using magnetic
multipoles.  Other trapping geometries including Halbach arrays~\cite{blumler2023practical} and Ioffe traps~\cite{ioffe1965mirror} are also in common use as atom traps and could also be considered as elements of a future atomic tritium source.

\begin{figure}
\begin{centering}
\includegraphics[width=1\columnwidth]{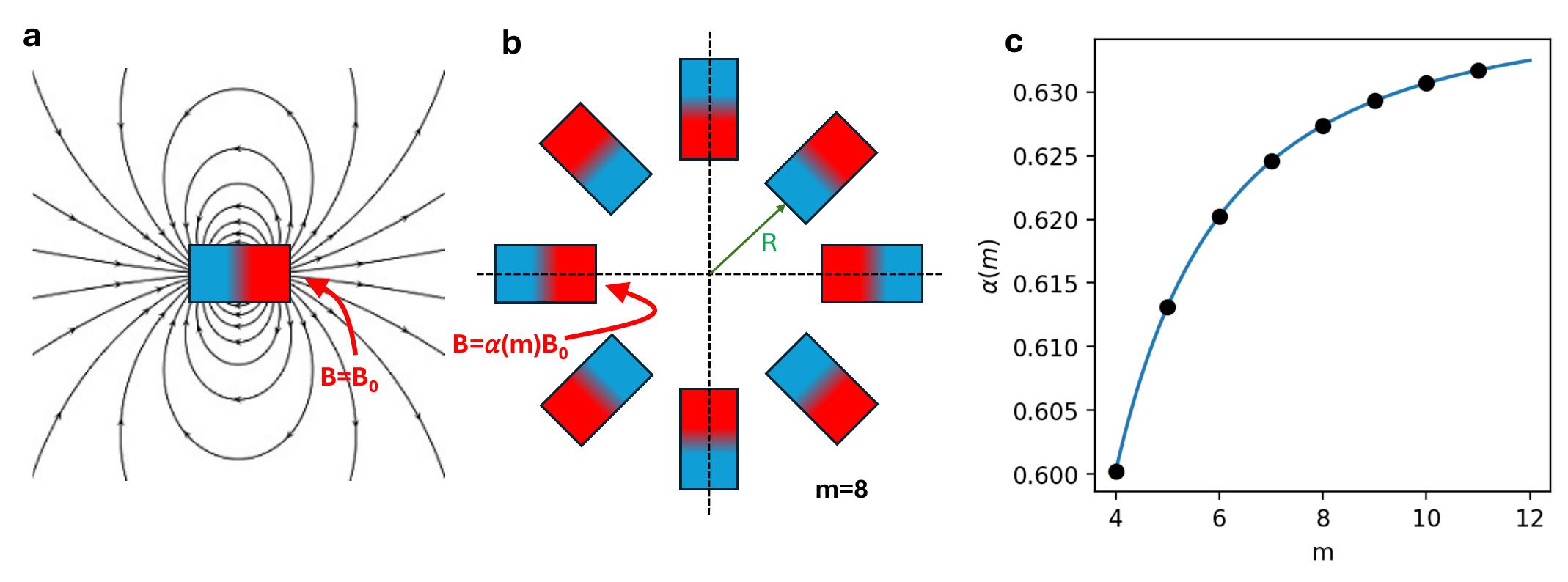}
\par\end{centering}
\caption{(a) Dipole field around a single magnet element. (b) Example
multipole configuration with $m=8$ elements. The average magnetic field
magnitude at R is given by $B=\alpha(m)B_{0}$. (c) Variation of
$\alpha(m)$ with~$m$.\label{fig:Left:-Dipole-field}}
\end{figure}

A multipole magnetic field configuration can be expressed analytically
as real and imaginary parts of the following complex expression \cite{wolski2011maxwell}
\begin{equation}
B_{x}+iB_{y}\propto\frac{(x+iy)^{\nu}}{R^{\nu}},
\end{equation}
where the normalization is fixed by specifying the magnetic field strength at reference radius $R$. A field that
is predominantly of multipole order $m$ can be generated by arranging
$m$ permanent magnets at a fixed radius and uniformly spaced angles,
with an alternating North-inward / South-inward ordering, as shown
in Fig.~\ref{fig:Left:-Dipole-field}. 
The average magnetic field
at the radius of the inner magnet surface in this configuration is
not exactly equivalent to the surface field of an individual magnet
in vacuum $B_0$, due to the segmentation of the geometry. If $m$ magnetic
dipole elements with inner and outer radius $R_{i}$ and $R_{o}$
respectively form the multipole, then the resulting $B$ field can
be shown to be \cite{blumler2023practical}
\begin{equation}
B_{x}(r)+iB_{y}(r)=B_{0}\frac{m^2\sin\left[(\frac{m}{2}+1)\pi/m\right]}{2\pi\left(\frac{m}{2}-1\right)\left(\frac{m}{2}+1\right)}\left(\frac{1}{R_{i}^{m/2-1}}-\frac{1}{R_{o}^{m/2-1}}\right)(x+iy)^{m/2-1}.
\end{equation}
For a thick magnet, $R_{o}\gg R_{i}\equiv R$ and we find that
\begin{equation}
B_{x}(r)+iB_{y}(r)=B_{0}\alpha(m)\left(\frac{x+iy}{R}\right)^{m/2-1},\quad\alpha(m)=\frac{\sin\left[(\frac{1}{2}+\frac{1}{m})\pi\right]}{2\pi\left(\frac{1}{4}-\frac{1}{m^{2}}\right)},
\end{equation}
from which it is apparent that $\nu=m/2-1$. The function $\alpha(m)$ is an order-1 number that encodes the effects
of segmentation of the magnet geometry, shown in Fig.~\ref{fig:Left:-Dipole-field}(c).

\begin{figure}
\begin{centering}
\includegraphics[width=1\columnwidth]{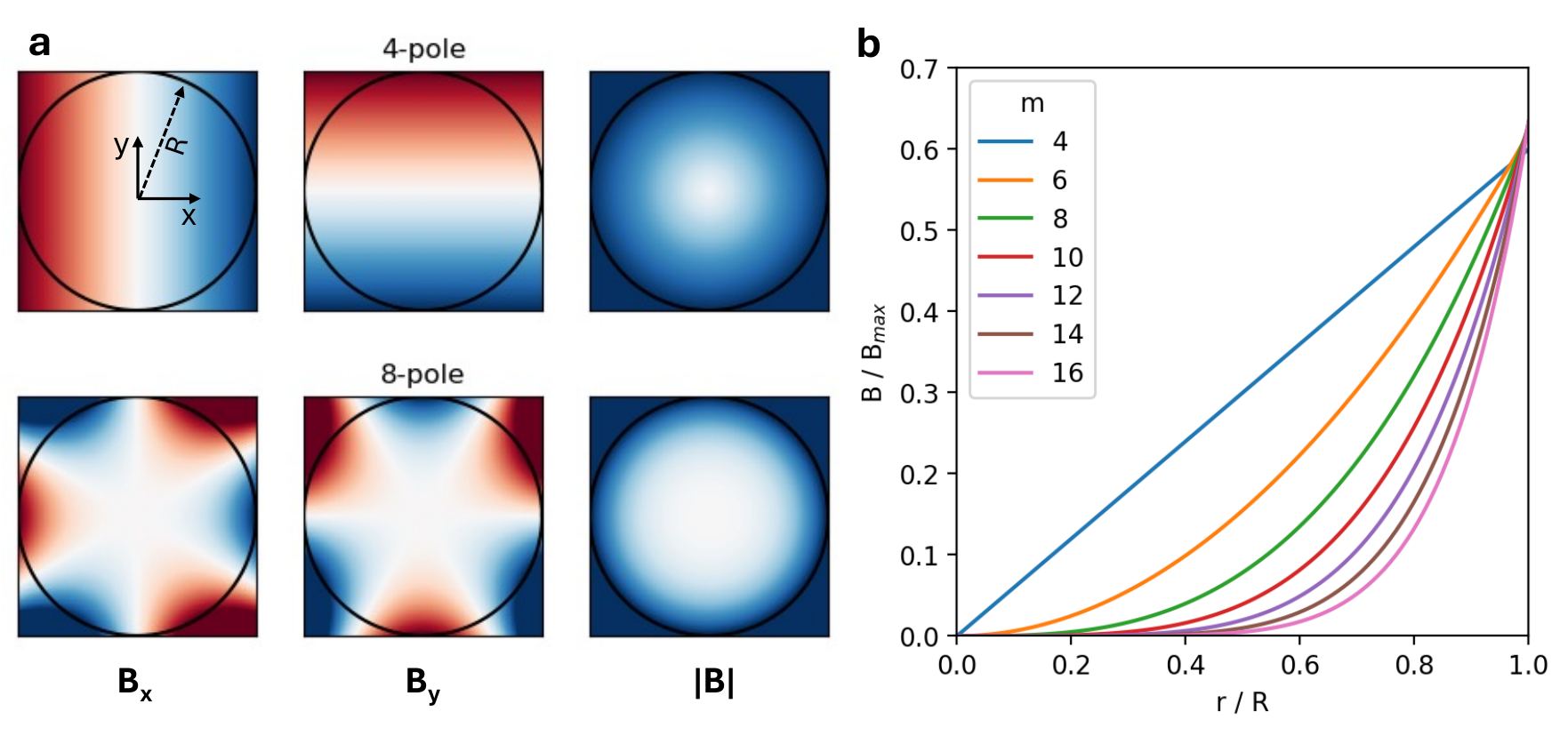}
\par\end{centering}
\caption{(a) $B$ fields in quadrupole and octupole guides, showing x and y
components and total $B$ field magnitude. (b) Effective trapping
magnetic field $B$ in a multipole guide, relative to the
surface field of a single magnet element $B_{0}$.\label{fig:Left:-B-fields}}
\end{figure}

Fig.~\ref{fig:Left:-B-fields}(a) shows geometrical multipole
field maps for a quadrupole and octupole guide. 
Fig.~\ref{fig:Left:-B-fields}(b) shows the radial dependence
of the field strength, scaled to the surface field of the magnet,
accounting for the finite segmentation correction $\alpha(m)$. We now proceed to analyze the trapping and cooling in multipole guides analytically. 

\begin{figure}
\begin{centering}
\includegraphics[width=1\columnwidth]{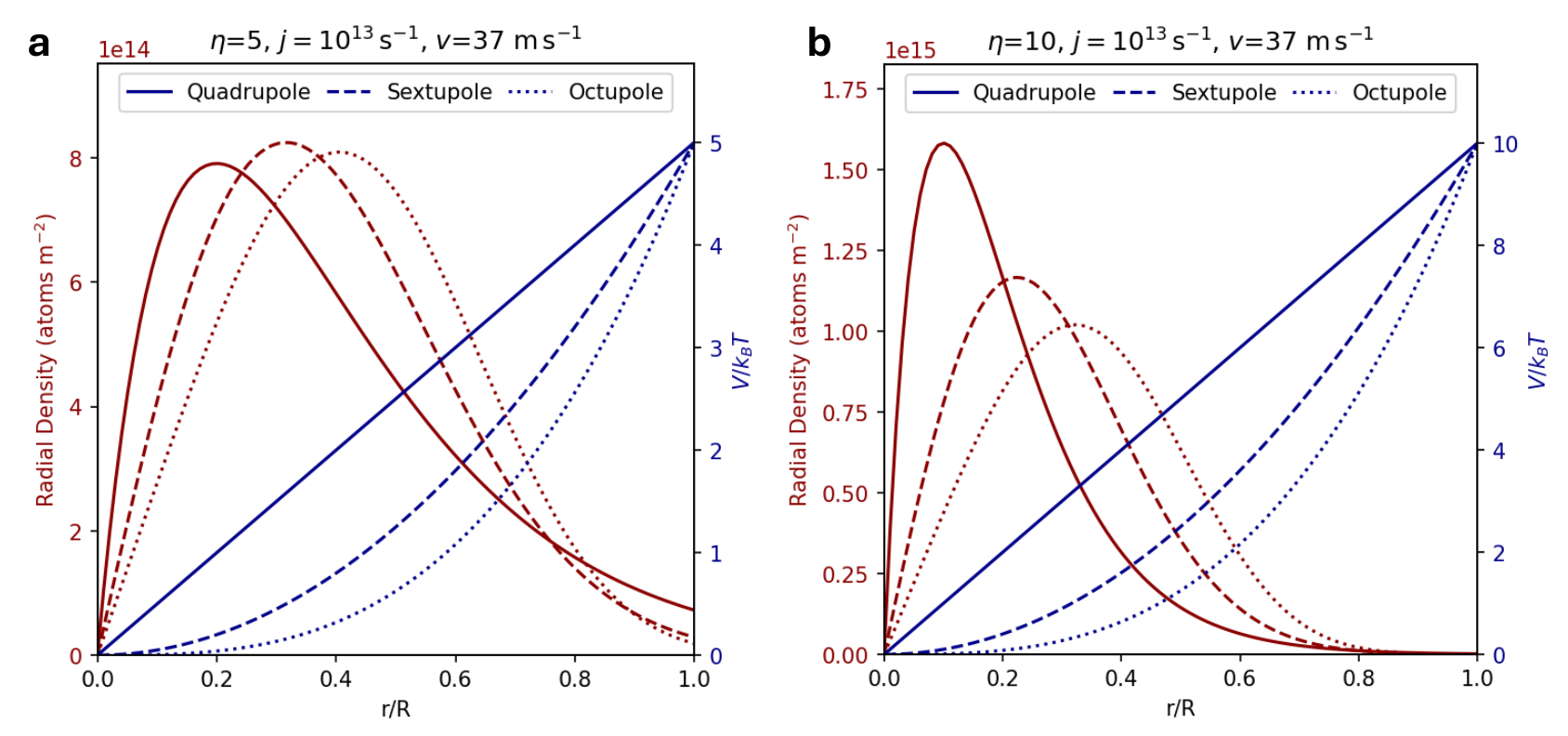}
\par\end{centering}
\caption{Radial density maps of particles in multipole guides compared to their
potential curves. (a) Shows barrier height $\eta=5$ and (b) shows
$\eta=10$, evaluated at representative system parameters. The blue lines show the potential shapes and the red show radial density distributions of atoms thermalized into these potentials. \label{fig:Radial-density-maps}}
\end{figure}

The trapping potential depends on
the magnitude  $B_0$ of the magnetic field,
\begin{equation}
V(r)=\alpha\mu_a B_{0}\left(\frac{r}{R}\right)^{\frac{m}{2}-1}.
\end{equation}
Low-field seeking atoms will tend to thermalize toward the center
of the guide because this is where $|B|$ is the smallest. Furthermore, the larger the multipolarity, the flatter the trap and lower the field gradient near the center. For a fixed individual magnet element surface
strength $B_{0}$, a higher multipolarity will hold a given number
of atoms at a lower average density. This has implications for the rate of collisions, which will be a relevant
input to the definition of the cooling geometry. If atoms with mass $M$ are allowed
to thermalize into a linear multipole guide and the trapping potential
is large enough to effectively trap the atoms (i.e. $V(R)\gg k_{B}T$),
they will become distributed in position $\vec{x}$ and momentum $\vec{p}$ according
to the Boltzmann distribution,

\begin{equation}
f(\vec{x},\vec{p})=D\left(\frac{1}{2Mk_{B}T}\right)^{3/2}\exp\left[-\frac{\left(p_x\right)^{2}+\left(p_y\right)^{2}+\left(p_z-p_{z}^{0}\right)^{2}}{2Mk_{B}T}-\frac{V(\vec{x})}{k_{B}T}\right],
\label{eq:DistFunc}
\end{equation}
where $D$ is a normalization constant that fixes the total number of
atoms and T is the temperature. This expression allows for a mean transverse momentum $p_{z}^{0}$
along the multipole, since the atoms are not trapped in this direction,
and the density at any position $\vec{x}$ can be obtained by integrating
the distribution function $f$ over the momentum degrees of freedom,

\begin{equation}
\rho(\vec{x})=\int d^{3}p\,f(\vec{x},\vec{p}).
\end{equation}
Eq.~\ref{eq:DistFunc} has the notable property that the momentum
and position dependencies are factorizable, which means that in an
equilibrated beam the velocity and position distributions are decoupled, simplifying calculations. For
a steady state beam that is moving along $\hat{z}$, we can express
the normalization factor $D$ in terms of the beam particle throughput,
by noting that if there is a bulk velocity $v=p_{z}^{0}/M$ then the
current $j(z)$ of atoms through a given $z$ surface $S$ is related
to the particle density $\rho(\vec{x})$ as
\begin{equation}
j(z)= v\int_{S} dxdy\,\rho(\vec{x}),
\end{equation}
which gives $D$ in terms of $j$ and $v$, with the result that
\begin{equation}
f(\vec{x},\vec{p})=j\frac{\eta^{\frac{4}{m-2}}}{\pi v\Gamma\left[\frac{m+2}{m-2}\right]R^{2}}\left(\frac{1}{2\pi Mk_{B}T}\right)^{3/2}\exp\left[-\frac{\left(p_x\right)^{2}+\left(p_y\right)^{2}+\left(p_z-p_{z}^{0}\right)^{2}}{2Mk_{B}T}-\eta\left(\frac{r}{R}\right)^{\frac{m}{2}-1}\right].\label{eq:FulLMB}
\end{equation}
We have introduced above the barrier height parameter $\eta$, which encodes
the depth of the trapping field in units of $k_{B}T$,
\begin{equation}
\eta\equiv\frac{\mu_a B_{\rm max}}{k_{B}T}=\alpha(m)\frac{\mu_a B_{0}}{k_{B}T},
\end{equation}
where $B_{\rm max}$ is the maximum value of the magnetic `wall' field encountered by atoms.  Fig.~\ref{fig:Radial-density-maps} shows plots at
fixed $R=1$ cm of the radial density ($2\pi r\rho(r)$) of atoms in an equilibrated
system compared against the radial potential curve for some representative
cooling parameters and two values of evaporative barrier height $\eta$.

\subsection{Analytic treatment of evaporative cooling in a guide \label{subsec:Cooling-dynamics}}

Atoms will evaporate if their transverse kinetic energy exceeds the magnetic potential at the wall.   As atoms evaporate, the remaining trapped atoms will, through collisions, equilibrate to a new temperature.     We now discuss the properties of this evaporation process on thermodynamic grounds.


A gas of $N$ atoms trapped in a 3D well has a total energy made up of the kinetic energy $\frac{3}{2}k_BTN$ and the potential energy $\langle V\rangle$ which depends on the shape of the potential. The virial theorem relates these two quantities.  If the potential is of the form $V\propto r^\nu$, then the total potential energy is:
\begin{eqnarray}
\left<V_{\rm}\right>&=&\frac{2}{\nu}\left<\frac{d}{2}k_BTN\right>,
\end{eqnarray}
where $d$ is the dimensionality of the well, and the operation $\langle \rangle$ indicates averaging over all particles at a given time.  

In a seminal paper~\cite{hess1986evaporative}, Hess advanced an analytic model to illustrate cooling by considering the evolution of kinetic and potential energies within an evaporating ensemble.  The Hess model is in fact incomplete, since it neglects the density of energy states in the trap, but it is informative of the basic principles of evaporative cooling and becomes relatively accurate in the limit $m,\nu\rightarrow\infty$. In this case, the rate of change of energy of the gas is determined by the evaporative cooling power $\dot{Q}_{\rm ev}$
\begin{eqnarray}
\frac{3}{2}k_B\frac{d}{dt}(TN)=\dot{Q}_{\rm ev},
\end{eqnarray}
where the evaporative cooling power depends on the threshold $\eta$ via
\begin{eqnarray}
\dot{Q}_{\rm ev}&=&(\eta+\eta')k_BT\dot{N}.
\end{eqnarray}
Here $\eta'$ is an order-one number characterizing the mean excess energy of an atom escaping over the potential barrier.   Hess~\cite{hess1986evaporative} suggests  $\eta'=2$ per particle. A simple calculation shows that for a flat trap (corresponding to the limit $\nu\rightarrow \infty$), $\eta'=3/2$. Other values of $\nu$ lead to different numerical results for $\eta$, with a significant impact on cooling rates.

In addition to evaporation, other loss terms arise from spin-flip and dipolar losses~\cite{stoof1988spin,zygelman2010electronic}, and from scattering off atoms in the background gas.  Our quantitative estimates suggest that these are significantly sub-leading, and as such only evaporation is considered in our analysis.  One is also free to interchange kinetic and potential energy components without loss by suitable variations of the magnetic confinement.  Such a term can be explicitly included when part of the cooling is achieved by isentropic expansion via reduction of the magnetic field~\cite{hess1986evaporative}.    

When the gas is not at rest as in Hess's analysis but is collectively moving along a magnetic guide, the kinetic energy of the bulk motion adds new terms to the above equations.  This additional kinetic energy is not in equilibrium with the internal temperature of the gas because in a magnetic guide there is no viscosity or friction (unless special measures are taken to introduce it).  If the mean bulk velocity is $\langle{v}\rangle$ the kinetic energy per atom due to bulk motion is $K\equiv\frac{1}{2} M\langle{v}\rangle^2$. Each atom that evaporates takes this additional kinetic energy with it, in addition to its heat.  The evaporative cooling power becomes
 \begin{eqnarray}
\dot{Q}_{\rm ev}&=&(\eta+\eta')k_BT\dot{N}+K\dot{N}.
\end{eqnarray}
The energy flow equation becomes:
\begin{eqnarray}
\frac{3}{2}(T\dot{N}+N\dot{T})+\frac{1}{k_B}K\dot{N}&=&(\eta+\eta')T\dot{N}+\frac{1}{k_B}K\dot{N}. \label{eq:coolnotslow}
\end{eqnarray}
If there is no coupling between the axial and radial degrees of freedom other than that arising from the collisions between the atoms, it is not possible to reduce the mean longitudinal speed of the beam.  As such, the bulk kinetic energy cancels and plays no role in cooling through evaporation.  Instead we find
\begin{equation}
\left(\frac{3}{2}-\eta-\eta'\right)\frac{\dot{N}}{N}=-\frac{3}{2}\frac{\dot{T}}{T}.
\end{equation}
Setting $\eta'=\frac{3}{2}$, as expected for $\nu\rightarrow\infty$,
\begin{equation}
\frac{d\ln{T}}{d\ln{N}}\equiv\gamma=\frac{2}{3}\eta.\label{eq:gammanoslow}
\end{equation}
The dimensionless number $\gamma$ is termed the cooling exponent.  Integrating,
\begin{eqnarray}
    N&\propto&T^{1/\gamma}, \label{eq:particlesvstemp}
\end{eqnarray}
where the constant of proportionality is determined by initial conditions.

\begin{figure}[t]
    \centering
    \includegraphics[width=0.75\linewidth]{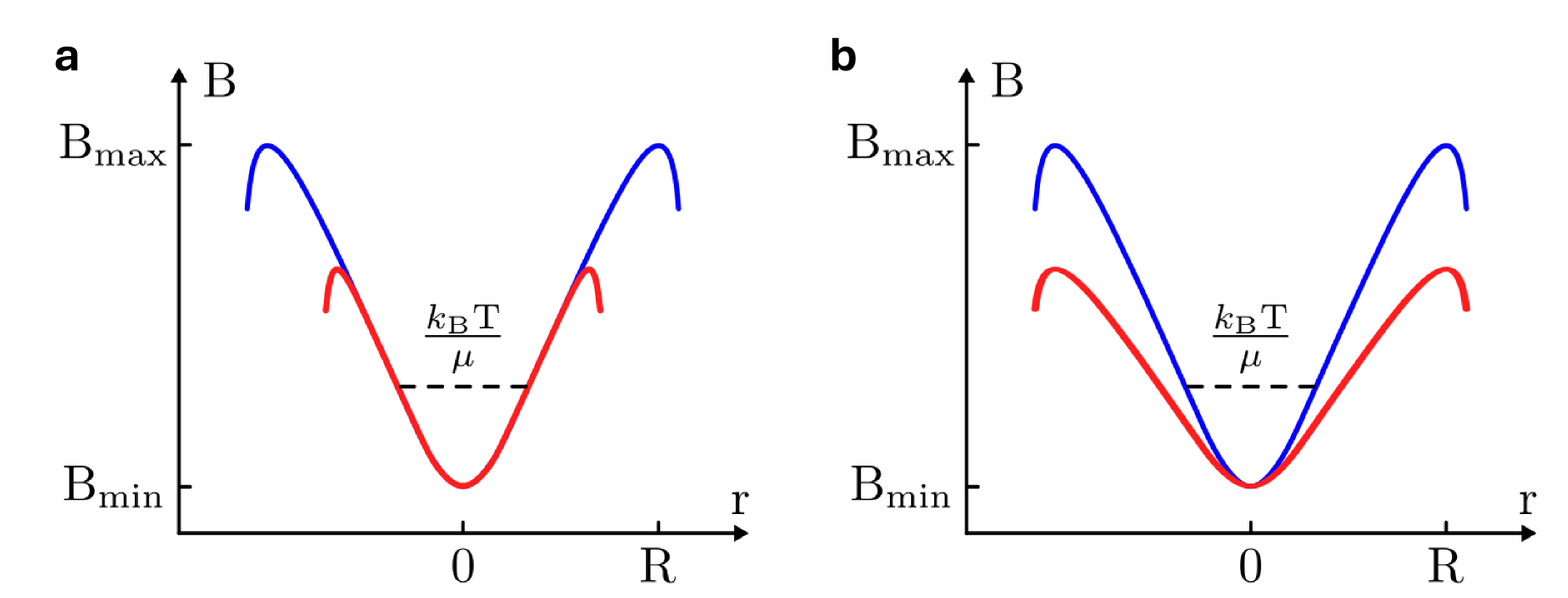}
    \caption{Two strategies for reducing the height of the magnetic wall in the guide to follow the reduction in beam temperature. In (a) the wall height is lowered while the guide trapping gradient is unchanged, whereas in (b) both wall height and trap shape are adjusted.  In both cases, the change between blue and red lines indicates the change in potential shape from a more upstream  location (blue) to a more downstream one (red).}
    \label{fig:reductionstrategies}
\end{figure}

To achieve steady cooling, the magnetic guide `wall' height as the gas cools needs to be lowered to maintain the value of $\eta$ near its optimum.  Choosing the magnetic geometry such that $\eta$ is constant imposes that $B_{\rm max}$ and $T$ remain proportional. Figure~\ref{fig:reductionstrategies} shows two strategies from a continuum of possibilities.
In Fig.~5(a) the guide potential is unchanged and atoms settle into ever lower potentials as they cool.  It has the property of shrinking the beam diameter steadily, in contrast to 5(b) where the field is reduced but the size of the magnetic guide is the same. 

A key principle omitted from the analysis of Hess is that the energy distribution of atoms thermalized into a potential well will have a non-trivial density of states in total energy $E$, deviating from a Maxwell Boltzmann distribution by a factor $g(E)/\sqrt{E}$.  An analytic model of the evaporative cooling process that includes this effect and hence is applicable for finite $\nu$ was provided by
Davis \etal~\cite{davis1995analytical}. Under this model, a trapping potential is applied
that scales like $V(r)\propto r^{\nu}$ along $d$ of the axes, with
no trapping in $3-d$ of them. In such a trap, the fraction of particles
with energy below some cut $\eta$ is
\begin{equation}
f_{N}^{<\eta}=\frac{1}{N}\int_{0}^{\eta k_B T}dE\,g(E)e^{-(E-\tilde{\mu})/k_{B}T}.
\end{equation}
For an $r^{\nu}$
trapping potential $g(E)$ is, following from Eq.~\ref{eq:FulLMB}, 
\begin{equation}
g(E)\sim\sqrt{E-\beta r^{\nu}},
\end{equation}
with $\beta$ a constant. The chemical potential $\tilde{\mu}$ fixes the normalization
of the distribution, and we can solve for it through applying the
constraint,
\begin{equation}
1=\frac{1}{N}\int_{0}^{\infty}dE\,g(E)e^{-(E-\tilde{\mu})/k_{B}T}.
\end{equation}

Finding $\tilde{\mu}$ in this way and then re-expressing the energy in terms
of dimensionless variable $\epsilon=E/k_{B}T$ we find the following
expression for the fraction of the particles or energy that lie above or below $\eta$,
\begin{eqnarray}
f_{N}^{>\eta}&=\frac{\int_{\eta}^{\infty}d\epsilon\,\left(\epsilon^{\frac{1}{2}+\frac{d}{\nu}}e^{-\epsilon}\right)}{\int_{0}^{\infty}d\epsilon\,\left(\epsilon^{\frac{1}{2}+\frac{d}{\nu}}e^{-\epsilon}\right)}&=\Gamma\left[\frac{3}{2}+\frac{d}{\nu},\eta\right]/\Gamma\left[\frac{3}{2}+\frac{d}{\nu}\right],\quad\quad f^{<\eta}_{N}=1-f^{>\eta}_{N}\\
f_{E}^{>\eta}&=\frac{\int_{\eta}^{\infty}d\epsilon\,\epsilon\,\epsilon^{\frac{1}{2}+\frac{d}{\nu}}e^{-\epsilon}}{\int_{0}^{\infty}d\epsilon\,\epsilon\,\epsilon^{\frac{1}{2}+\frac{d}{\nu}}e^{-\epsilon}}&=\Gamma\left[\frac{5}{2}+\frac{d}{\nu},\eta\right]/\Gamma\left[\frac{5}{2}+\frac{d}{\nu}\right],\quad\quad f^{<\eta}_{E}=1-f^{>\eta}_{E}
\end{eqnarray}
where $\Gamma$ with one or two arguments represents the complete or incomplete gamma function, respectively. The mean energy lost per particle leaving the trap (which we have
called $\eta'$) can be found by calculating the weighted energy of
the particles with $\epsilon>\eta$, which is given in terms of the ratio
of two incomplete gamma functions by
\begin{equation}
\eta'=\frac{\int_{\eta}^{\infty}d\epsilon\,\left(\epsilon^{\frac{1}{2}+\frac{d}{\nu}+1}e^{-\epsilon}\right)}{\int_{\eta}^{\infty}d\epsilon\,\left(\epsilon^{\frac{1}{2}+\frac{d}{\nu}}e^{-\epsilon}\right)}-\eta=\frac{\Gamma\left[\frac{5}{2}+\frac{d}{\nu},\eta\right]}{\Gamma\left[\frac{3}{2}+\frac{d}{\nu},\eta\right]}-\eta.
\end{equation}
To find the cooling exponent, we need the fraction of particles
$f_{N}$ and energy $f_{E}$ below energy cut $\eta$, and then
we can obtain $\gamma$ via
\begin{equation}
\frac{d\ln T}{d\ln N}\equiv\gamma=\frac{f_{E}^{<\eta}-f_{N}^{<\eta}}{f_{N}^{<\eta}-1}.
\end{equation}
We thus obtain an expression for $\gamma$ that accounts for the trap density
of states for finite multipolarity $\nu$,
\begin{equation}
\gamma=\frac{\Gamma\left[\frac{5}{2}+\frac{d}{\nu},\eta\right]/\Gamma\left[\frac{5}{2}+\frac{d}{\nu}\right]}{\Gamma\left[\frac{3}{2}+\frac{d}{\nu},\eta\right]/\Gamma\left[\frac{3}{2}+\frac{d}{\nu}\right]}-1.
\end{equation}
This is notably different from the value of $\gamma$ provided in
the original work of Hess~\cite{hess1986evaporative}, which is, in this notation,
\begin{equation}
\gamma_{\rm Hess}=\frac{2}{3}\left(\eta+\eta'-\frac{3}{2}-\frac{d}{\nu}\right).
\end{equation}

\begin{figure}
\begin{centering}
\includegraphics[width=0.49\columnwidth]{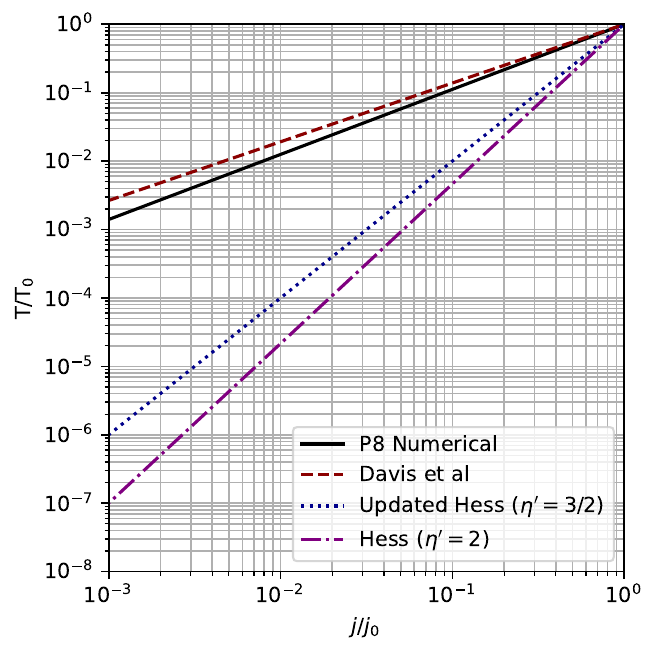}
\par\end{centering}
\caption{Comparison of temperature reduction $T/T_0$ vs atom transport efficiency $j/j_0$ for a quadrupole with $\eta=5$ under various models. The lines correspond to the analytic models of Hess~\cite{hess1986evaporative} with both Hess's proposed $\eta'=2$ and with $\eta'=\frac{3}{2}$ as expected for a cut Maxwell Boltzmann distribution; the Davis {\em et al.}~\cite{davis1995analytical} model that includes the density-of-states effect; and the numerical model discussed in this paper.  The reported behavior is universal, though the rate at which these temperature and current reductions occur carry additional dependencies including atomic mass and cross section.\label{fig:KetterleAndHess1}}
\end{figure}

\begin{figure}
\begin{centering}
\includegraphics[width=0.49\columnwidth]{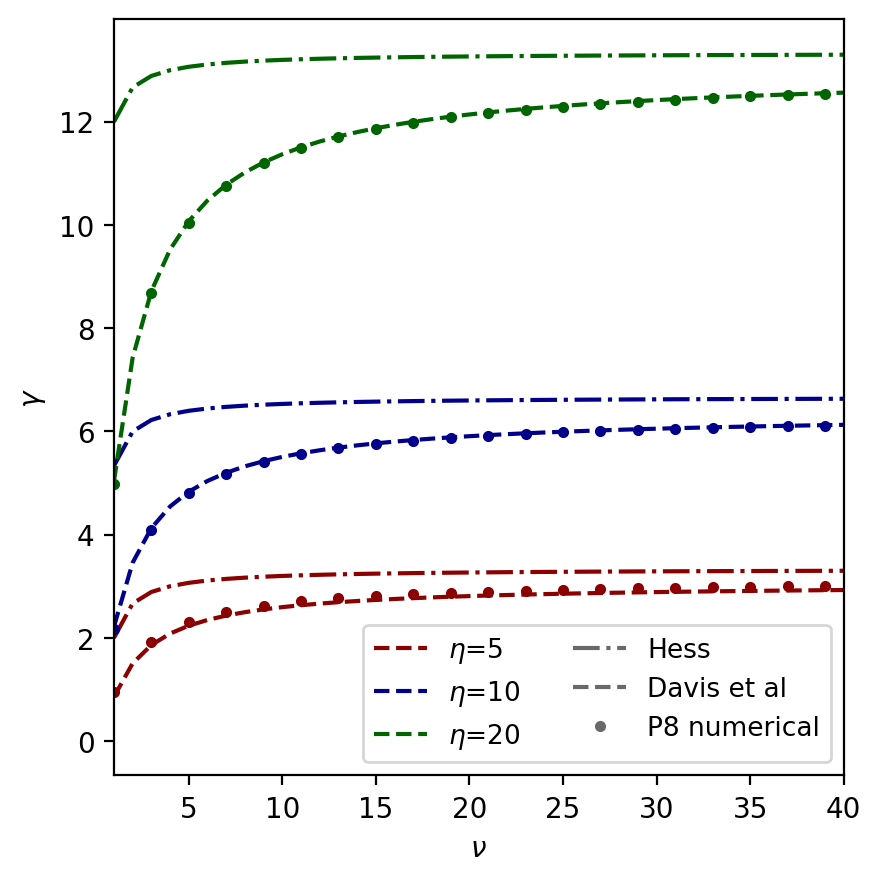}
\par\end{centering}
\caption{Cooling exponent $\gamma$ as a function of multipolarity $\nu$ for several values of evaporation cut $\eta$ in various models.  A good agreement is observed between the Davis {\em et al.} model~\cite{davis1995analytical} and the numerical calculations presented in this work. \label{fig:KetterleAndHess2}}
\end{figure}

In the case of fixed guide geometry and evaporation cut $\eta$, the achievable temperature reduction $T/T_0$ for a given atom transport efficiency $j/j_0\sim N/N_0$ (with initial values of temperature $T_0$ and current $j_0$) scales as a power law with exponent $\gamma$, per Eq.~\ref{eq:particlesvstemp}.   Figure~\ref{fig:KetterleAndHess1} compares the various $\gamma$ predictions discussed above, for
a finite quadrupole with $\eta=5$ to the numerical model advanced in this work. Figure~\ref{fig:KetterleAndHess2} compares the Hess and Davis models for various values of the multipolarity and also compares against the numerical model that will be presented in subsequent sections.  In all models, we observe the general trend that cooling efficiency for a specific temperature drop is better for larger multipolarity $\nu$ and for larger evaporation cut $\eta$. Both figures show good agreement between the numerical solutions of this work and the Davis model which incorporates the density-of-states effect. At the multipolarities of interest to this study
the Hess approximation provides an imperfect prediction, but as the multipolarity becomes large the Hess
approximation becomes increasingly accurate. This correspondence occurs
because at large $\nu$ the trap becomes flatter. In this case, the density
of states approaches that for the free-space Maxwell Boltzmann expectation
$g(E)\sim\sqrt{E}$, over most of the volume of the trap.

We turn now to the problem of slowing the longitudinal component of the bulk motion, under the simple flat-potential $\nu\rightarrow\infty$ Hess model.  We consider slowing the beam by transferring longitudinal energy to transverse for evaporative cooling.  The transfer is effected by introducing magnetic perturbations along the beamline to mimic a viscous or ``friction'' force.  The additional transverse energy is evaporated away, if the beam is dense enough to regain thermal equilibrium.  The perturbations can be quasi-continuous (the beam is consistently slowed down during cooling) or discrete (the beam is slowed in steps between evaporative cooling segments).
 
We first consider the quasi-continuous case.  The energy flow becomes
 \begin{eqnarray}
 \frac{3}{2}k_B\frac{d}{dt}(TN) + \dot{N}K+N\dot{K}&=&\dot{Q}_{\rm ev}.
 \label{eq:coollongitudinal}
 \end{eqnarray}
 We introduce the ``mass flow number'',  which is the ratio of the bulk transport velocity to the mean thermal speed in the co-moving frame
 \begin{equation}
 \mathcal{M}\equiv\sqrt{\frac{2K}{3k_BT}}=\frac{\langle v\rangle}{\sqrt{\langle v^2 \rangle- \langle v \rangle^2}},
 \end{equation}
 which is related to the ``Mach number'' $\tilde{\mathcal{M}}$ via  $\tilde{\mathcal{M}}=\sqrt{\frac{3}{\tilde{\gamma}}}\mathcal{M}$, with the adiabatic constant $\tilde{\gamma}=5/3$ for an ideal monatomic gas (though notably, spin-polarized Li and T do not necessarily exhibit the thermophysical properties of a simple monatomic gas due to the long-range triplet potentials between atoms~\cite{sandouqa2018thermophysical}). $K$ may be written conveniently in terms of $\mathcal{M}$:
 \begin{eqnarray}
     K&=&\frac{3}{2}\mathcal{M}^2k_BT.
 \end{eqnarray}
 We assume that the mass flow number $\mathcal{M}$,   
 in the quasi-continuous scenario, is held constant as the beam is cooled. 
 The evaporative cooling power is
\begin{eqnarray}
\dot{Q}_{\rm ev}&=&(\eta+\eta')k_BT\dot{N}+\frac{3}{2}\mathcal{M}^2k_BT\dot{N}.
\end{eqnarray}
Then,
 \begin{equation}
 \left(\frac{3}{2}-\eta-\eta'\right)T\dot{N}=-\left(\frac{3}{2}+ \frac{3\mathcal{M}^2}{2}\right)N\dot{T},
 \end{equation}
 leading to cooling exponent
 \begin{equation}
 \gamma=\frac{\frac{2}{3}\left(\eta+\eta'\right)-1}{{\cal M}^{2}+1}. \label{eq:coolslowgamma}
 \end{equation}
 To avoid excessive beam losses we require $\gamma\gtrsim 1$, a condition that favors low $\cal{M}$ and high $\eta$.  On the other hand, larger values of $\cal{M}$ are better for assuring atoms flow forward in the beam, and $\eta$ cannot be too large or the cooling process will become very slow and the beamline prohibitively long.  The smallest value of $\mathcal{M}$ in directed flow is for effusive escape from an orifice, for which the mean longitudinal velocity is 
\begin{eqnarray}
\langle v\rangle&=&\sqrt{\frac{\pi k_BT}{2M}},
\end{eqnarray}
and hence 
\begin{eqnarray}
\mathcal{M}^2\geq \mathcal{M}^2({\rm effusion})&=&\frac{\pi}{6}.
\end{eqnarray}
The optimal choice of $\eta$ is fixed by the allowed beamline length, which determines the highest achievable efficiency.  

An alternative approach, using discrete-perturbations to accomplish slowing of the beam turns out to be more efficient (less costly in evaporated atoms).  In this approach atoms keep the same longitudinal speed while undergoing transverse cooling, and then encounter a slowing perturbation in the multipole field that reheats the beam and slows the bulk motion.  The process is repeated until the desired temperature is reached.  

The transverse cooling takes place as before, though since $\dot{K}=0$ in the cooling segments, the $\mathcal{M}$ term is absent in the cooling exponent.  The cooling segments thus change $N$ and $T$ but not $K$, reducing the spread of the velocity distribution without affecting its mean value.  The slowing perturbations transfer kinetic energy into heat, with an amount of reheating equal to the change in kinetic energy $K$.  The slowing segments thus change $T$ and $K$ but not $N$, converting longitudinal energy into internal energy of the beam.  The beam parameters $(N,K,T)$ at three points along the axis of each section are hence as follows: $(N_1,K_1,T_1)$ gives the particle number, bulk kinetic energy per particle, and temperature on entering a transverse-cooling section; $(N_2,K_2,T_2)$ gives those parameters at the end of the cooling segment and before the slowing perturbation; and $(N_3,K_3,T_3)$ gives the parameters after the slowing perturbation and at the entrance to the next cooling segment. In the cooling segment, $\gamma$ is as given in Eq.~\ref{eq:gammanoslow}, whereas in the slowing segment $N$ is unchanged,
\begin{eqnarray}
 \frac{N_2}{N_1}&=&\left(\frac{T_2}{T_1}\right)^{1/\gamma},\quad\quad\frac{N_3}{N_2}=1\label{eq:particlefraction}
\end{eqnarray}
We assume the slowing perturbation can be tuned so that the mass flow number entering each cooling segment is always the same (with $\mathcal{M}$ representing the value at both points 1 and 3).  Energy conservation implies that the reheating effect on the beam is 
\begin{equation}
    T_3-T_2=-\frac{2}{3}\frac{K_3-K_2}{k_B},\quad\quad K_1=K_2     
\end{equation}    
such that
\begin{equation}
    \frac{T_3}{T_1}=\frac{1}{1+\mathcal{M}^2}\left[\left(\frac{N_3}{N_1}\right)^{\gamma}+\mathcal{M}^2\right]. \label{eq:beamheating}
\end{equation}
Applying Eqs.~\ref{eq:particlefraction} and \ref{eq:beamheating} repeatedly in a cascade of cooling and slowing steps, one can arrive at a final temperature, beam velocity, and particle number. Examples of such cooling trajectories are later calculated and presented in Figs.~\ref{fig:Cooler-with-no}, \ref{fig:Tritium-system-performance}.  

The models discussed in this section provide some indications of  the relevant scaling laws for some performance parameters in a MECB system.  Even in the most advanced models discussed, however,  no information about times and distances can be obtained without appending a detailed treatment that involves cross sections, beam dimensions, currents, and other inputs.  We now proceed to develop this more complete treatment.  Several parameters will present themselves for optimization, and we explore these in the following sections.

\subsection{Triplet scattering cross sections \label{subsec:Triplet-scattering-cross}}

\begin{figure}[t]
\begin{centering}
\begin{minipage}[b][0.6\columnwidth][c]{0.6\columnwidth}%
\begin{center}
\includegraphics[width=1\columnwidth]{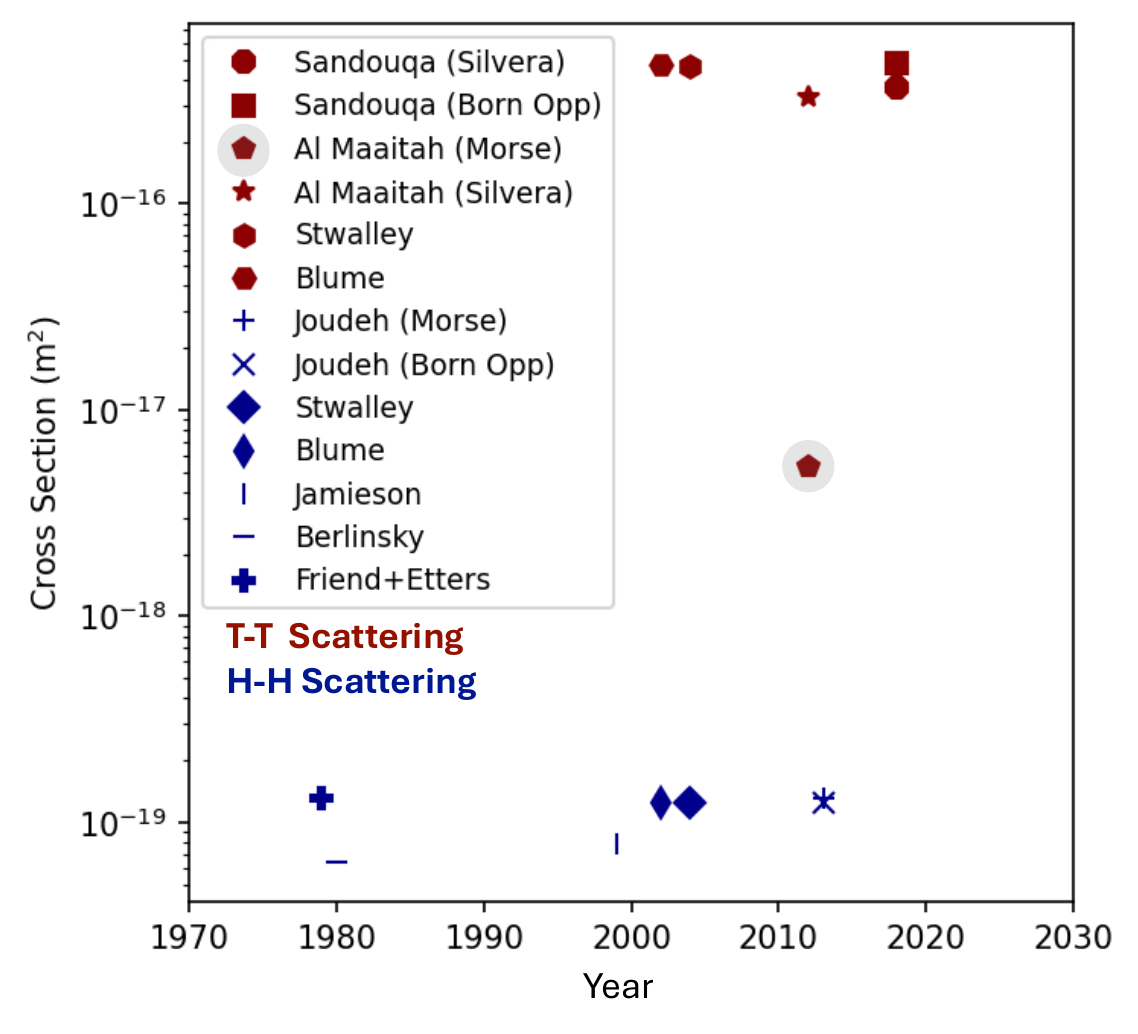}
\par\end{center}%
\end{minipage}%
\begin{minipage}[b][0.6\columnwidth][c]{0.4\columnwidth}%
\begin{center}
\begin{tabular}{|c|c|c|}
\hline 
Species & Triplet s-wave $\tilde{a}$ & Cross Section\tabularnewline
\hline 
\hline 
H & 70 pm & $1.2\times10^{-19}\mathrm{m}^{2}$\tabularnewline
\hline 
T & -4200 pm & $4.4\times10^{-16}\mathrm{m}^{2}$\tabularnewline
\hline 
$^{6}$Li & -11900 pm & $3.5\times10^{-13}\mathrm{m}^{2}$\tabularnewline
\hline 
$^{7}$Li & -144 pm & $5.3\times10^{-19}\mathrm{m^{2}}$\tabularnewline
\hline 
\end{tabular}
\par\end{center}%
\end{minipage}
\par\end{centering}
\caption{Left: Triplet s-wave scattering cross sections for T (red) and hydrogen (blue) from existing literature.  Right: Tabulated reference cross section values for H and Li isotopes that are used in Project 8 atomic cooling calculations, including those presented in this paper.\label{fig:Triplet-S-wave-scattering}}
\end{figure}

For a thermalized distribution in a potential well, individual atoms
are constantly interacting with one another, randomly sharing their
energy to maintain a Maxwell Boltzmann distribution. The
frequency of collisions that eject atoms from the guide and lead to
cooling, as well as those that redistribute
energy to re-thermalize the distribution, is dictated by the local
density of atoms at each position in the guide and the triplet
$(|\uparrow_e\rangle|\uparrow_e\rangle\rightarrow|\uparrow_e\rangle|\uparrow_e\rangle)$
scattering cross section $\sigma$.   Because T atoms are indistinguishable bosons, any combination of low-field-seeking hyperfine states will undergo s-wave triplet elastic scattering. For low energy collisions of identical particles, $\sigma$
can be expressed exclusively in terms of the triplet s-wave scattering length
via $\sigma=8\pi \tilde{a}^{2}$. The
scattering length $\tilde{a}$ can, in turn, be calculated from interatomic potentials.
A popular potential for atomic H and T calculations
is the Silvera potential~\cite{silvera1980solid}, which is derived
from a multi-parameter fit to numerical solutions of the Schr\"{o}dinger
equation for pairs of atoms at fixed separation that was evaluated
in Ref.~\cite{kolos1975improved}. The Morse~\cite{dugan1973ground}
and Born Oppenheimer \cite{jamieson1999collisions,blume2002formation}
potentials provide alternatives.

Solution of the Schr\"{o}dinger equation in the presence of these potentials
provides the s-wave phase shift, and hence the scattering length.
The challenge of these calculations lies in the fact that the scattering
length is very sensitive to the repulsive core of the potential, and
small changes to this shape can have a dramatic impact on the final
result. Calculations of the H-H scattering length
have been performed by Berlinsky \cite{berlinsky1980elementary}, Friend and Etters \cite{friend1980dilute}, Jamieson \etal~\cite{jamieson1999collisions}, Blume \etal~\cite{blume2002formation}, Stwalley
\cite{stwalley2004collisions}, and Joudeh \cite{joudeh2013scattering}, 
with a consensus value
of  $\approx0.7$ \r{A} in the low temperature limit, where the relative momentum in the collisions tends toward zero. T-T
scattering length calculations have been presented by  Blume \cite{blume2002formation}, Stwalley \cite{stwalley2004collisions}, Al Maaitah \cite{al2012scattering}, and Sandouqa and Joudeh
\cite{sandouqa2018thermophysical}, and
suggest a much larger value of $\approx42$ \r{A}, corresponding to a scattering
cross section nearly four orders of magnitude larger for T-T than for H-H.
For the purposes of evaporative cooling, this increased scattering
length is highly beneficial as it results in much faster thermalization
of the interacting atomic vapor, and relaxes the beam density requirement.
Fig.~\ref{fig:Triplet-S-wave-scattering} shows the calculated s-wave
cross sections collected from the aforementioned literature, for H-H
and T-T scattering. Apart from one outlier highlighted in grey (T-T scattering using the Morse potential), the literature provides a consistent picture of the expected triplet scattering lengths relevant for evaporative cooling.  The most modern calculations for the singlet scattering length of atomic H include nonadiabatic effects and QED corrections~\cite{lai2022precision,lai2025precision}, which can have effects at the tens-of-percent level. Calculations of the triplet scattering length of tritium or hydrogen that include these effects do not appear to be available at the present time.

The triplet scattering cross sections for $^{6}$Li have been
calculated in Ref.~\cite{abraham1997triplet} (-$2160\pm250\,a_{0}$)
and Ref.~\cite{o2000optical} (-2240 $a_{0}$) based on potentials
from Refs.~\cite{konowalow1979electronic,konowalow1984most,schmidt1985ground,yan1996variational,zemke1993analysis,abraham1997triplet}.

For present purposes we take values for the relevant cross
sections to be those tabulated in Fig.~\ref{fig:Triplet-S-wave-scattering},
right, which are representative of the available theoretical literature,
with the s-wave triplet cross section of $^{6}$Li having $\sigma\approx  3.5\times10^{-13}$~m$^{2}$
and T having $\sigma\approx4.4\times10^{-16}\,\mathrm{m}^{2}$. The additional complication associated with the vanishing triplet scattering cross section for identical fermions in the case of $^6$Li is not prohibitive in this geometry, as the travel time in the beam is insufficient for significant relaxation of the non-stretched states to occur. The Li beam will instead be comprised of a mixture of the low-field-seeking hyperfine states, thus maintaining the large triplet scattering cross section even at very low temperature. The increased cross section of  Li  over  T  implies
that the Li demonstration apparatus can operate at somewhat lower current
and hence lower instantaneous beam density than the T apparatus,
making it possible to explore these techniques with currently available
oven-based sources and Zeeman slowing systems. 

\subsection{Beam cooling via evaporation \label{subsec:Beam-cooling-via}}

\begin{figure}
\begin{centering}
\includegraphics[width=.99\columnwidth]{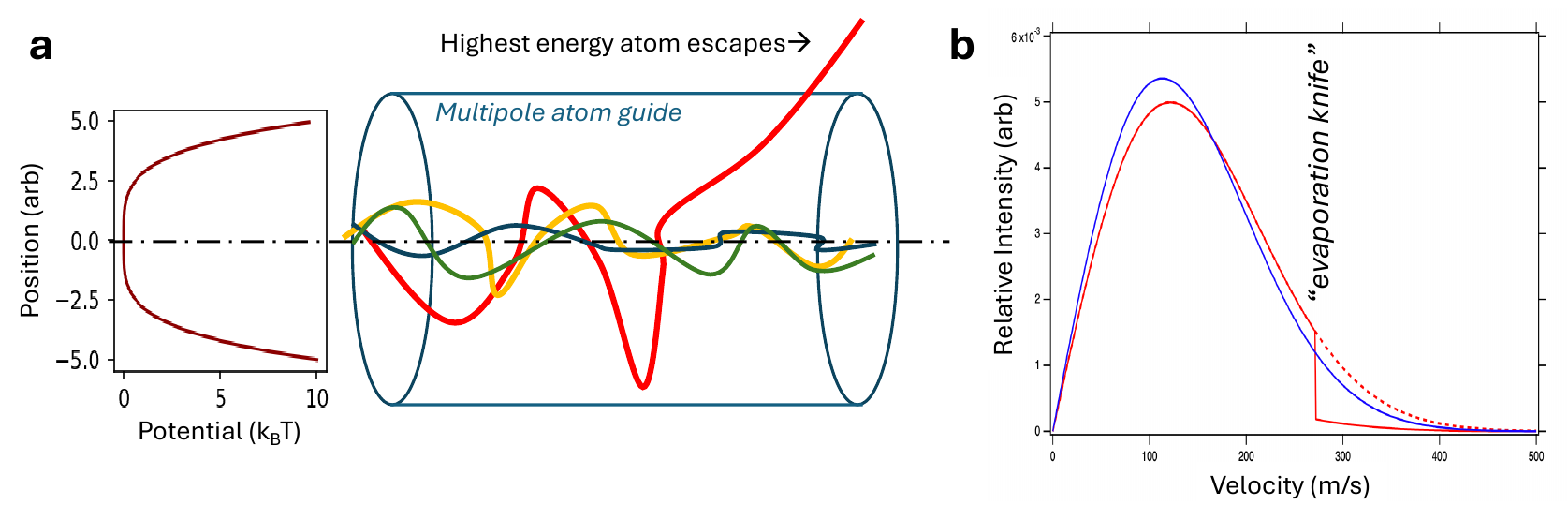}
\par\end{centering}
\caption{(a) Diagram illustrating the dynamics of evaporative cooling from
a multipole guide. The highest energy sub-population with energies above the ``evaporation knife'' leave, and the
remaining particles thermalize to a lower energy Maxwell-Boltzmann
distribution. (b) A cut 1-K Maxwell-Boltzmann distribution (red) and the
lower temperature distribution (blue) toward which this ensemble will evolve upon
rethermalization. The temperature has decreased 12\% and the intensity by 7\%. \label{fig:Top:-Diagram-illustrating}}

\end{figure}

Evaporation occurs as a result of particles randomly receiving an
upward fluctuation in energy following a collision, that leads them
to have enough transverse momentum to escape from the guide.  
Both the shape of the potential and a cartoon illustrating the evaporation dynamics are  shown schematically in Fig.~\ref{fig:Top:-Diagram-illustrating}(a).

Any trapped gas will eventually attain a local Maxwell Boltzmann distribution through thermalizing collisions, and an evaporation cut can be applied to the high tail of this distribution. The highest energy subset of particles will then escape, leaving only a small tail of still-escaping atoms above the energy cut (indicated by the red curve in Fig.~\ref{fig:Top:-Diagram-illustrating}(b)).  The remaining particles redistribute energy through further collisions, leading
to a lower temperature Maxwell Boltzmann distribution (the blue curve in Fig.~\ref{fig:Top:-Diagram-illustrating}(b)). Since evaporation is driven by the subset of collisions that populate high tail whereas all collisions contribute to thermalization, the Maxwell Boltzmann form is recovered quickly.  The thermalizing collisions
also redistribute energy between the $p_{x}$, $p_{y}$ and $p_{z}$ directions,
which implies that the temperature in the longitudinal and transverse
directions (quantified via the spread of the momentum phase space
distribution) is always isotropic. The collisions conserve both
total energy and total momentum, so while the distribution is cooled
to a smaller phase space volume, its mean momentum remains constant
(assuming there is no preferential evaporation direction, valid in the
the case of a radial evaporation cut). A thermalized
beam in a linear multipole guide will thus maintain a mean momentum
in the $z$ direction as it cools. This implies that
the challenge of producing a beam to load a stationary trap has two
components: cooling, which involves reducing the width of the thermalized
momentum phase-space distribution; and slowing, which involves removing
the mean momentum. Both of these goals can be accomplished in static
magnetic configurations, and in this section and the next we outline the
methods to calculate their effectiveness in representative
geometries.

\begin{figure}
\begin{centering}
\includegraphics[width=0.9\columnwidth]{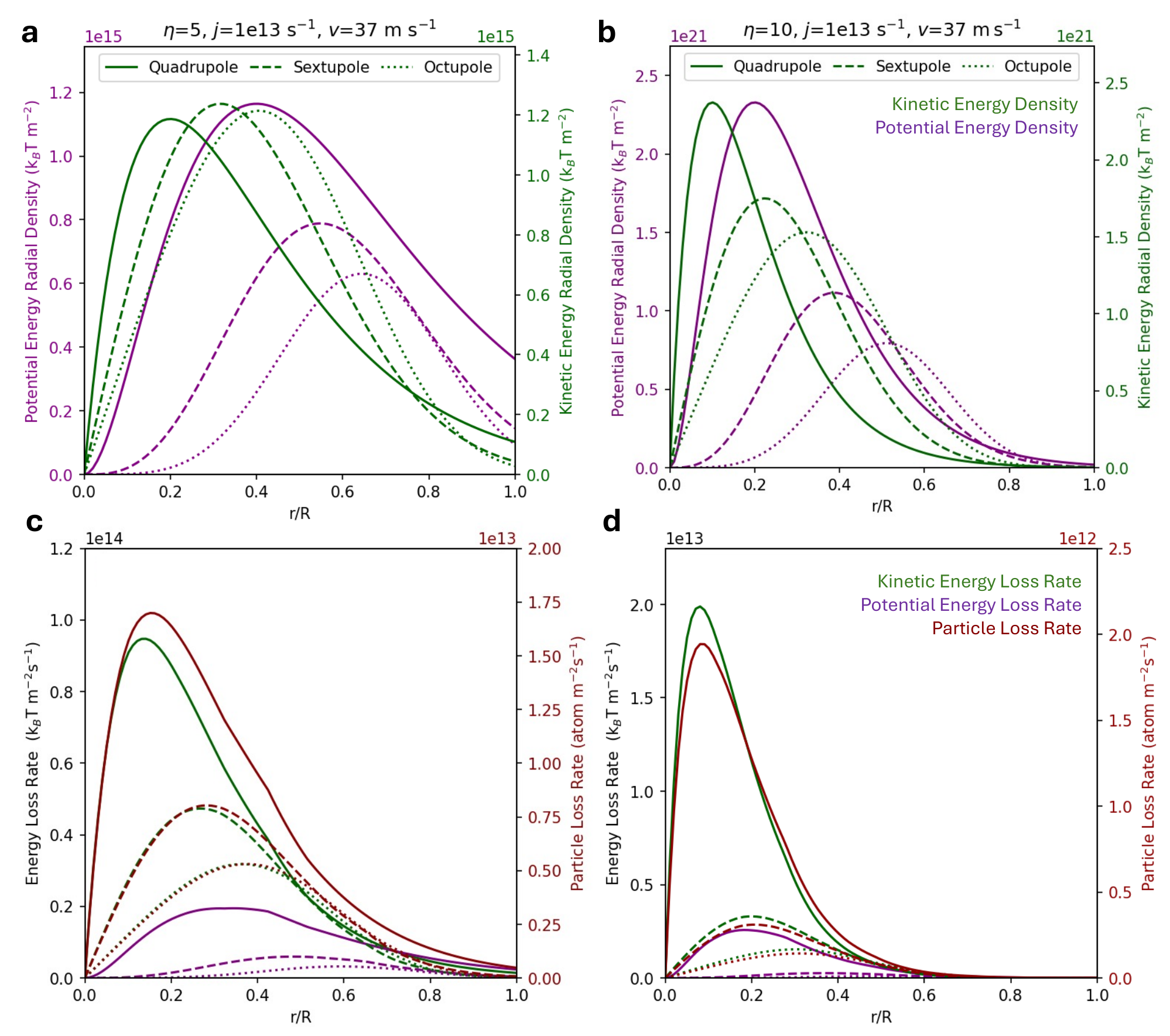}
\par\end{centering}
\caption{Radial kinetic and potential energy distributions in various multipole
guides for two different evaporation cuts, $\eta=5$ left and $\eta=10$
right, for a representative set of Li parameters.  Panels (a) and (b) show particle and energy radial densities per unit length of guide; (c) and (d) show the loss rates of particles and energy.
\label{fig:FigKinPot}}
\end{figure}

The equipartition theorem states that for a well-thermalized distribution,
the linear energy density in a given $z$-slice of the guide of width
$\delta z$ will be
\begin{equation}
E(z)=\int dxdy\,\left({\cal E}_{kin}+{\cal E}_{pot}\right)\delta z=\left(\frac{3}{2}+\frac{4}{m-2}\right)N(z)k_{B}T.\label{eq:EnergyZ}
\end{equation}
Here $N(z)=\int dxdy\,\rho(z)\delta z$ is the number of particles
between $z$ and $z+\delta z$. The kinetic and potential energy distributions
can be evaluated for a thermalized distribution, as
\begin{equation}
{\cal E}_{kin}=\frac{3}{2}k_{B}T\rho,\quad{\cal E}_{pot}=V(r)\rho.
\end{equation}
Fig.~\ref{fig:FigKinPot} shows examples of the kinetic and potential
energy distributions across the beam for several multipole
configurations. 
We are interested in the rates of change of particle number and energy with $z$, which we denote as $\Gamma_X$. For a beam moving with bulk velocity $\langle v\rangle$ (which hereafter we will denote simply as $v$, for brevity), these can be connected to the time derivatives in the beam co-moving rest frame via
\begin{equation}
\Gamma_{X}\equiv\frac{1}{X}\frac{dX}{dz}\sim\frac{1}{v}\frac{1}{X}\frac{dX}{dt}.
\end{equation}
As a result of Eq \ref{eq:EnergyZ}, the change in
temperature as a function of longitudinal position along the beam
can be found in terms of the losses in energy and particle number
due to evaporation, as
\begin{equation}
\Gamma_{T}=\Gamma_{E}-\Gamma_{N}.\label{eq:TempEvol}
\end{equation}
To obtain $\Gamma_{E}$ and $\Gamma_{N}$ we must find the evaporation
rates, integrated over the spatial density distribution in the guide.
A particle will leave if its transverse momentum following a collision
is more than the trap depth, when accounting also for potential energy
$V(r)$ at the location of the collision. The ejection threshold $\eta_{cut}$ and critical momentum $p_{cut}$
for loss of particles at each position is thus given by
\begin{equation}
\eta_{cut}\equiv\frac{p_{cut}^{2}(\vec{x})}{2Mk_{B}T}=\eta\left[1-\left(\frac{r}{R}\right)^{\nu}\right].
\end{equation}

The rate of particles acquiring momenta above this value from collisions
is obtained from the time dependence of the phase space distribution
$\frac{df}{dt}$ at each position. In a generic out-of-thermal-equilibrium
system this time dependence can be obtained from the Boltzmann transport
equation,
\begin{equation}
\frac{\partial}{\partial t}f(\vec{x},\vec{p})=-\frac{\vec{p}}{M}\cdot\nabla_{x}f(\vec{x},\vec{p})+\nabla_{x}V(\vec{x})\cdot\nabla_{p}f(\vec{r},\vec{p})+{\cal I}(\vec{r},\vec{p}),
\end{equation}
whose terms represent, from left to right: time evolution of phase
space, mass flow due to inertia, force from applied potential, and
collision integral $\mathcal{I}$ that leads to thermalization. For spherical collisions,
${\cal I}$ has the form
\begin{equation}
{\cal I}(\vec{r},\vec{p}_{4})=\frac{\sigma}{\pi M^{2}}\int d^{3}p_{3}\left[\int d^{3}p_{1}d^{3}p_{2}\right]\delta_{p}^{3}\delta_{E}\left\{ f(\vec{r},\vec{p}_{1})f(\vec{r},\vec{p}_{2})-f(\vec{r},\vec{p}_{3})f(\vec{r},\vec{p}_{4})\right\}.
\end{equation}
The delta functions $\delta_{p}^{3}\delta_{E}$ (defined in Appendix A) enforce energy
conservation in two-body collisions, and indices 1,2 label the initial state and 3,4 the final state particles. To quantify evaporation rates, the relevant contribution to
the phase space distribution is that associated with the collision
integral $\left[\frac{\partial f}{\partial t}\right]_{col}={\cal I}(\vec{x},\vec{p})$.
The relevant loss rates of position-dependent kinetic energy density,
potential energy density and particle density can be obtained
by numerical integration of the out-of-equilibrium phase space distribution
with an appropriate transverse momentum cut applied, as
\begin{equation}
\left[\frac{1}{{\cal E}_{pot}}\frac{d{\cal E}_{pot}}{dt}\right]_{p_{cut},\sigma,\rho,T}=\left[\frac{1}{\rho}\frac{d\rho}{dt}\right]_{p_{cut},\sigma,\rho,T}=\frac{\int_{p_{\perp}\geq p_{cut}}d^{3}p\left[\frac{\partial}{\partial t}f(p)\right]_{col}}{\int_{\infty}d^{3}pf(p)},
\end{equation}
\begin{equation}
\left[\frac{1}{{\cal E}_{kin}}\frac{d{\cal E}_{kin}}{dt}\right]_{p_{cut},\sigma,\rho,T}=\frac{\int_{p_{\perp}\geq p_{cut}}d^{3}p\frac{p^{2}}{2M}\left[\frac{\partial}{\partial t}f(p)\right]_{col}}{\int_{\infty}d^{3}p\frac{p^{2}}{2M}f(p)}.
\end{equation}
The time derivatives
$\left[\frac{\partial f}{\partial t}\right]_{col}$ appearing in the above equations encode the change in
the phase space distribution function due to redistribution of the
ensemble energy and momentum during evaporation. This time derivative
can be evaluated via the full collisional Boltzmann transport equation, as presented in  \hyperref[sec:AppendixA]{Appendix A}.

Our numerical simulations show that the calculated loss rates obey
very closely some expected simple scaling relations with temperature, cross
section and density, allowing them to be evaluated once for each value
of $\eta$ in conditions $\rho_{0},\sigma_{0}$ and $T_{0}$
and applied at generic $\rho,\sigma$ and $T$ via, for example,

\begin{equation}
\left[\frac{1}{{\cal E}_{pot}}\frac{d{\cal E}_{pot}}{dt}\right]_{p_{cut},\sigma,\rho,T}=\frac{\sigma}{\sigma_{0}}\frac{\rho}{\rho_{0}}\sqrt{\frac{T}{T_{0}}}\left[\frac{1}{{\cal E}_{pot}}\frac{d{\cal E}_{pot}}{dt}\right]_{p_{cut},\sigma_{0}.\rho_{0},T_{0}}.\label{eq:Scaling2}
\end{equation}

\begin{figure}
\begin{centering}
\includegraphics[width=0.6\columnwidth]{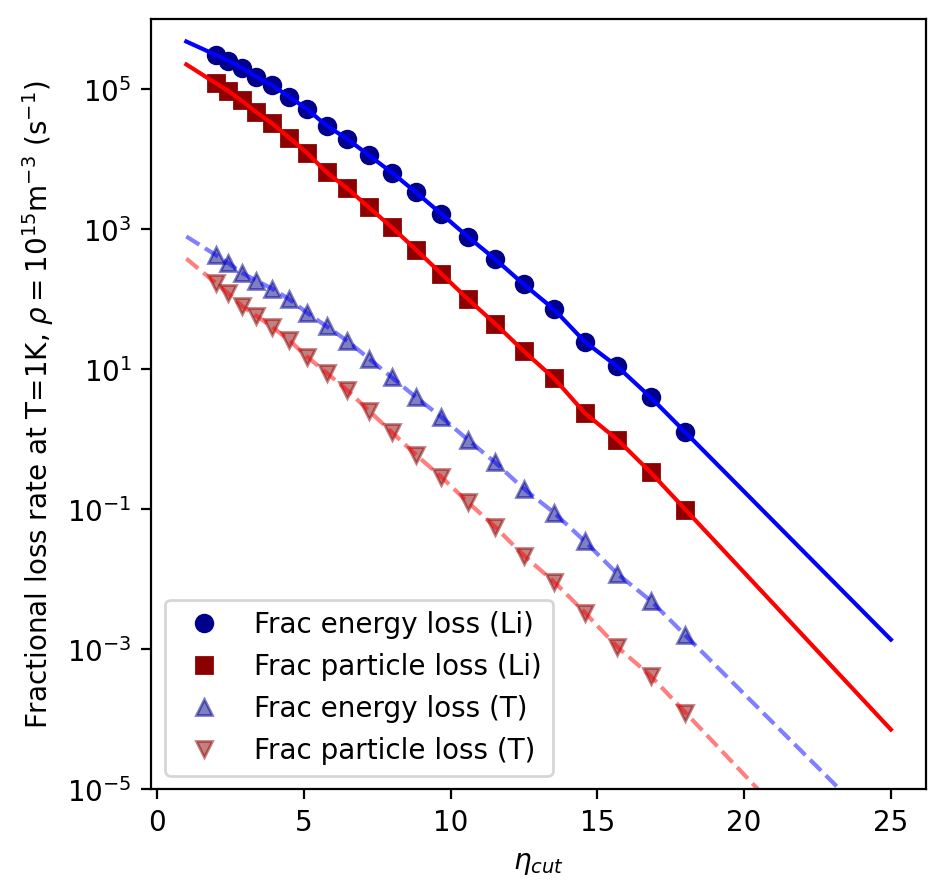}
\par\end{centering}
\caption{Fractional evaporation loss rates of energy (blue) and particles (red) as a function
of the effective evaporation cut. Note that the actual cut applied
varies as a function of position across the guide due to the different
amounts of potential energy stored in each particle undergoing an ejecting
collision. \label{fig:ThermCuts}}
\end{figure}

These quantities, evaluated in reference conditions, are shown in Fig.~\ref{fig:ThermCuts}, and used in the
calculations that follow via application of Eq.~\ref{eq:Scaling2}. The relative particle and energy loss rates are shown as red and blue lines respectively, in Fig.~\ref{fig:ThermCuts}.  Since only the highest energy particles leave, the relative energy loss rate exceeds the relative particle loss rate, leading to cooling.  The difference between these rates becomes larger for higher evaporation cut values, producing more efficient but slower cooling.  The
 longitudinal temperature evolution is found via Eq. \ref{eq:TempEvol},
with kinetic, potential and particle losses  integrated over
the transverse geometry of the guide.   These integrals take the form

\begin{figure}
\begin{centering}
\includegraphics[width=0.85\columnwidth]{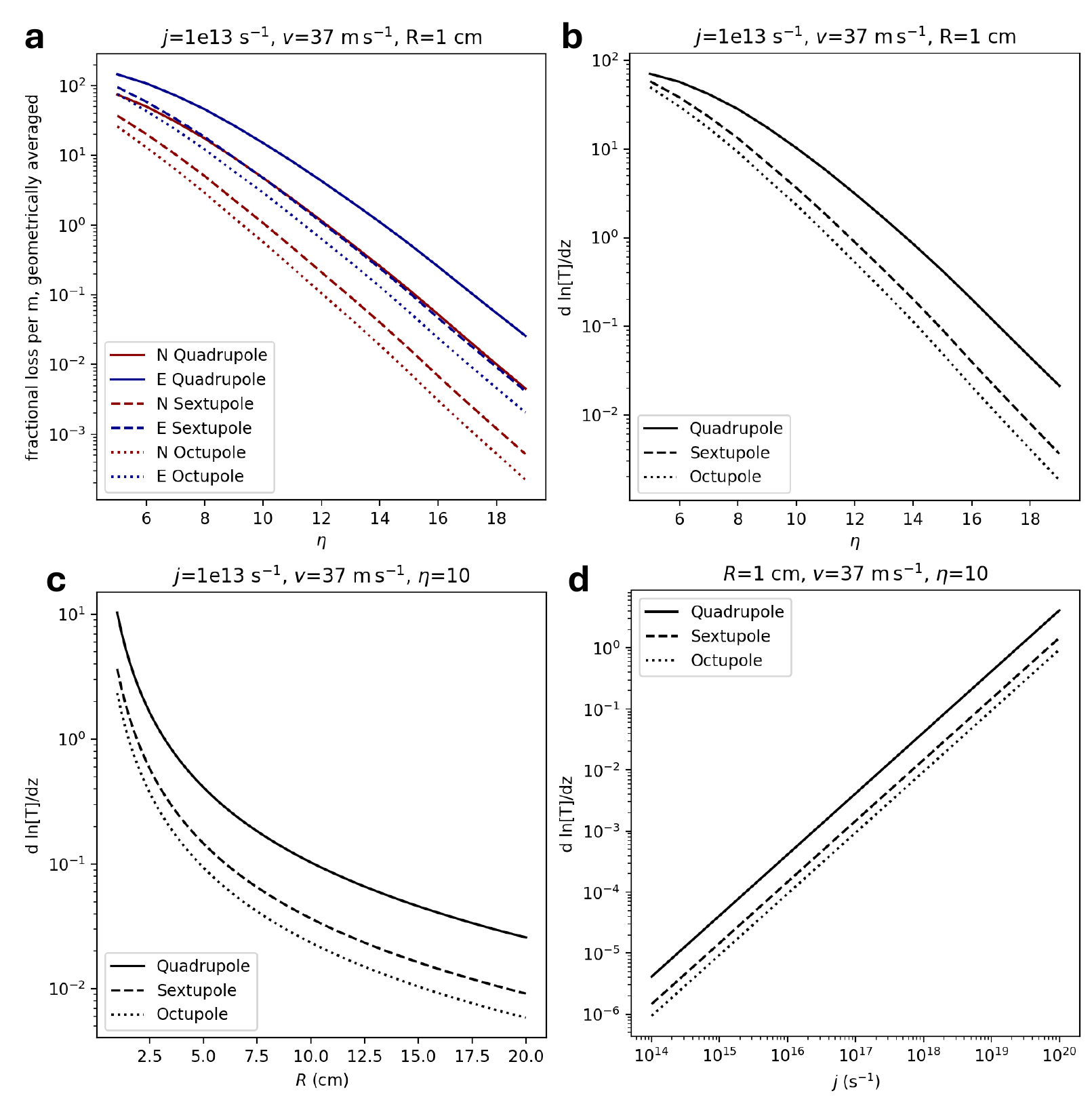}\caption{Cooling trajectories integrated over the guide. (a) dependence of energy and particle losses shown individually for different multipolarities and cooling cuts; (b) dependence of cooling exponent $\gamma$ on cooling cut with other parameters fixed; (c)  dependence of cooling exponent $\gamma$ on guide radius with other parameters fixed; (d) dependence of cooling exponent $\gamma$ on atomic current with other parameters fixed.\label{fig:Cooling-trajectories-integrated}}
\par\end{centering}
\end{figure}

\begin{equation}
\Gamma_{E}=\frac{\int dr\,r\rho(r)\left(\frac{3}{2}\left[\frac{1}{{\cal E}_{kin}}\frac{d{\cal E}_{kin}}{dz}\right]_{p_{cut}(r),\sigma,\rho,T}+V(r)\left[\frac{1}{{\cal \rho}(r)}\frac{d{\cal \rho}}{dz}\right]_{p_{cut}(r),\sigma,\rho,T}\right)}{v\left(\frac{3}{2}+\frac{4}{m-2}\right)N},\label{eq:GamE}
\end{equation}
\begin{equation}
\Gamma_{N}=\frac{\int dr\,r{\cal \rho}(r)\left[\frac{1}{{\cal \rho}(r)}\frac{d{\cal \rho}(r)}{dz}\right]_{p_{cut}(r),\sigma,\rho,T}}{v N}.\label{eq:GamN}
\end{equation}
Since the whole distribution is co-moving along the guide at velocity
$v$, the $z$ dependent loss rates can be found from these time dependent
rates via,
\begin{equation}
\Gamma_{k}\equiv \frac{1}{v}\left[\frac{1}{{\cal E}_{kin}}\frac{d{\cal E}_{kin}}{dt}\right]_{evap},\quad\Gamma_{N}\equiv \frac{1}{v}\left[\frac{1}{\rho}\frac{d\rho}{dt}\right]_{evap}.
\end{equation}
Examples of energy and particle loss rates as a function of 
potential well depth $\eta$ for various multipolarities is given
in Fig.~\ref{fig:Cooling-trajectories-integrated}(a), with
the consequent temperature evolution as a function of the input parameters $\eta,R,j$
shown in panels (b), (c), and (d) for each multipolarity.


For evaporation to be effective, the beam particles must maintain
thermal contact with one another. This implies a constraint on the
multipole radius $R$ and hence the beam radius, for a given throughput,
velocity and multipolarity. To obtain this constraint, consider that
the mean free path for a collision is given by
\begin{equation}
\lambda(r)=\frac{1}{\sqrt{2}\sigma\rho(r)}.
\end{equation}
where the factor of $\sqrt{2}$ in the denominator accounts for the fact that all of the relevant particles are in motion. Most of the particles are contained within a radius $R_{\rho}$ determined
by
\begin{equation}
R_{\rho}=R\eta^{-\frac{1}{\nu}}.
\end{equation}

In the regime $\lambda(r)\ll R_{\rho}$ we have many scatters per
crossing of the guide, which corresponds to a fluid-like flow; cooling
is unlikely to be efficient in such conditions, because particles
scattering above the evaporation cut cannot efficiently leave the
beam unless they happen to be on its periphery. On the other hand,
if $\lambda(r)\gg R_{\rho}$, the thermalization will be so slow that
evaporation will not effectively cool the distribution, since the highest energy particles will
leave the guide but remaining particles will not then be up-scattered to replace them.  In one pass through the beam, each atom traverses around $2R_{\rho}$ of path
length, and we introduce the dimensionless parameter $\zeta$ to quantify the opacity of the beam via $\zeta\equiv \frac{2 R_{\rho}}{\langle \lambda \rangle}$. The ideal regime is expected to be $\zeta\sim1$, and we  restrict our considerations to geometries that both
require physically achievable values of $R$ and also allow for sufficient
thermalization with $\zeta\gtrsim1$. 

\begin{equation}
\zeta=2\sqrt{2}\rho(0)\sigma R\eta^{-\frac{2}{m-2}}=\frac{2\sqrt{2}\eta^{\frac{2}{m-2}}\sigma j}{\pi v\Gamma\left[\frac{m+2}{m-2}\right]R},
\end{equation} 
where the on-axis density $\rho(0)$ is obtained from Eq.~\ref{eq:FulLMB}. Assuming $\zeta\sim1$ is maintained,
we now have have two dynamical variables, $j$ and $T$, whose
evolution is governed by two equations of motion,



 \begin{eqnarray} 
 \frac{d\ln T}{dz}&=&\frac{1}{v}\left[\frac{d\ln E}{dt}(T,j,R,B,v)-\frac{d\ln N}{dt}(T,j,R,B,v)\right],\\
 \frac{d\ln j}{dz}&=&\frac{1}{v}\left[\frac{d\ln N}{dt}(T,j,R,B,v)\right].
 \end{eqnarray}

Numerical solution of these equations using the inputs described above
provides cooling trajectories. Example trajectories for geometries
of interest will be explored in Sec. \ref{sec:MECB-Geometry-Proposal}.

We note that our treatment in this section assumes both that all particles scattering above the energy cut will leave the guide, and that there are not other loss mechanisms actively removing particles with a non-trivial energy bias during transport. Neither of these assumptions is precisely true in practice. Additional loss mechanisms beyond evaporation may include dipole losses, Majorana spin flips, and collisions with recombined tritium gas that can perturb the ideal dynamics discussed here.  These are all expected to be sub-leading effects, and we defer assessing their quantitative impact  to future work.

\subsection{Beam slowing via transverse perturbation \label{subsec:Beam-slowing-via}}

We now turn attention to slowing rather than cooling. In order to
remove the mean momentum from the beam,  some
form of friction against the guide must be introduced. Ignoring the effects of collisions,
the trajectories of individual particles in a guide obey a simple
equation of motion
\begin{equation}
\vec{F}=M\ddot{\vec{x}}=-\nabla V=-\mu_a\nabla|B|.
\end{equation}
For a straight multipole guide, the $B$ field magnitude is 
\begin{equation}
|B(\vec{r})|=\alpha(m)B_{0}\left(\frac{r}{R}\right)^{\frac{m}{2}-1}\left|\left(\begin{array}{c}
x/r\\
y/r\\
0
\end{array}\right)\right|.
\end{equation}
In this case, there is no force in the $\hat{z}$ direction and so
$p_{z}$ for each particle and hence for the whole ensemble is conserved
(this could also be considered to be a consequence of Noether's theorem,
given the symmetry of the potential in $\hat{z}$). In the special
case where $m=6$, {\em i.e.} for a sextupole guide, the $x$ and $y$ equations
of motion are also decoupled, whereas for other multipolarities
they become coupled.

If we introduce perturbations to the linear guide such that it no
longer has translational $\hat{z}$ symmetry, $p_{z}$ will no longer be conserved and an ensemble of particles will experience friction and become slowed.  In order to conserve energy this will generate heat, and this heat can be removed by further evaporation.
We aim to exploit this behavior to slow as well as cool the beam of
fast-moving atoms through evaporative cooling.

\begin{figure}
\begin{centering}
\includegraphics[width=0.93\columnwidth]{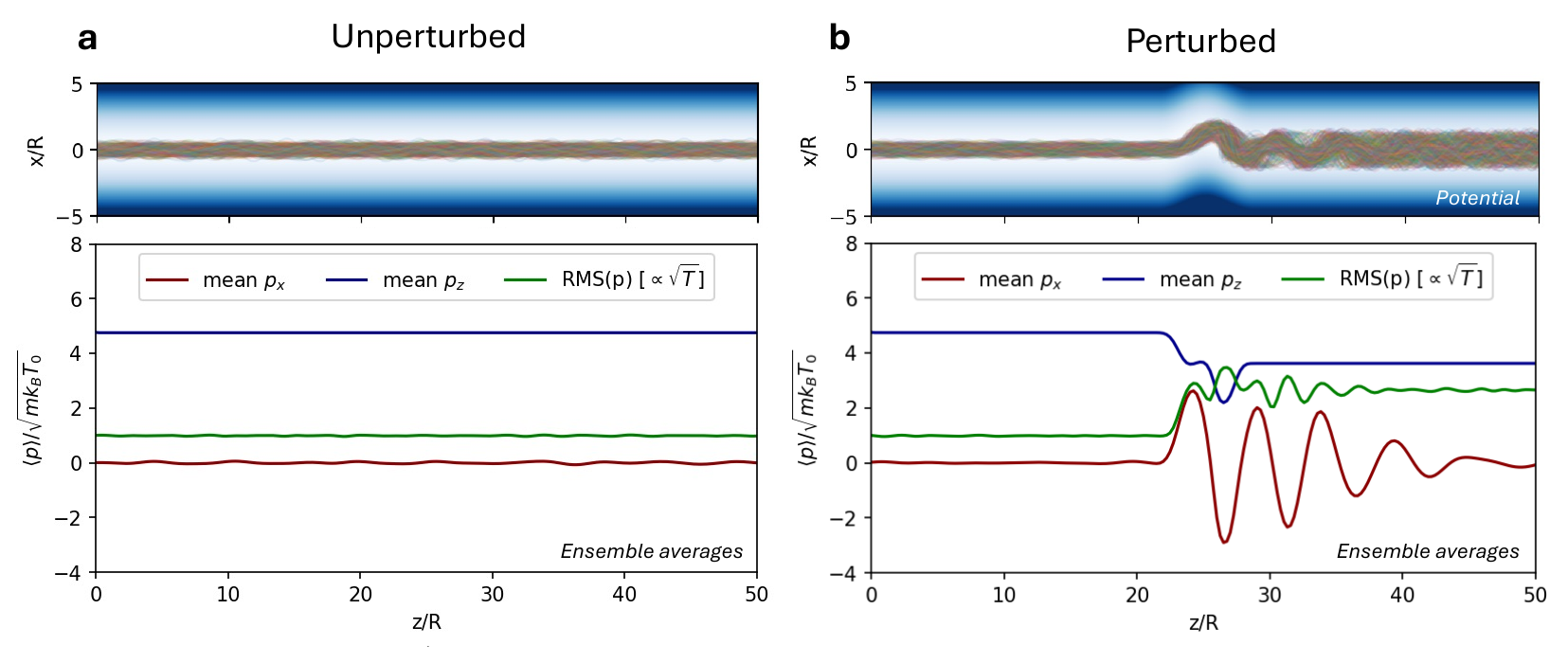}
\par\end{centering}
\caption{Effect of a perturbation generating friction that slows a fast-moving
beam. The top panels show the trajectories of particles moving in an applied magnetic potential, rendered in blue.  (a) Shows the dynamics of the unperturbed beam. (b) Shows a beam with a transverse perturbation at $z=25~R$, with $R$  the multipole radius. At the position of the perturbation
the mean momentum in $\hat{z}$ is reduced and the local temperature
is increased. The mean transverse momentum of the ensemble is initially
perturbed but eventually averages to zero as each particle emerges
from the kink with a different longitudinal wavelength. \label{fig:Effect-of-a}}
\end{figure}

A challenge in analyzing the effects of more complex magnetic geometries lies in defining the appropriate phase space distribution
 for particles entering them. If the guide has a non-trivial $\hat{z}$ profile
then we may no longer rely on our previous notion of a thermalized
co-moving ensemble, since the moving beam is explicitly
out of thermal equilibrium with the z-dependent geometry, and will
only achieve thermal equilibrium when it comes to rest with respect
to the magnetic potential. 

Our approach to this problem takes inspiration from scattering
theory. In a scattering problem, a well conditioned beam (of particles or waves) impinges
from spatial infinity on a localized potential, and the behavior of
the outgoing flux at infinity is studied. In analogy, for the present
problem we will consider a long segment of straight multipole
guide into which particles have thermalized as a co-moving distribution
from $z\rightarrow-\infty$; then they will encounter a localized
perturbation, and we will analyze the effect on the outgoing phase
space distribution as $z\rightarrow\pm\infty$.  This is the expected situation if the beam has reached equilibrium via collisions in a straight guide section before the perturbation.  A more complex geometry
can be constructed by applying a series of such perturbations between straight, thermalizing segments.   To ensure validity of the thermal incoming phase space distribution, any non-trivial kinks in the magnetic geometry should be separated by sufficient lengths of straight guide to allow for thermalization in between.
For the moment we will ignore the effect of collisions since the friction against the guide is approximately independent of how
strongly the beam is interacting. We will also restrict attention in
this section to harmonic guides, since this allows for analysis
in two dimensions rather than three.

As an example case, we will investigate scattering perturbations in
the guide of the form
\begin{equation}
V=\frac{c_{0}}{R^2}\left\{ \begin{array}{c}
x^2\\
\left(x-G(z)\right)^{2}\\
x^2
\end{array}\right.\begin{array}{c}
z<-a/2\\
-a/2<z<a/2\\
z>a/2
\end{array}
\end{equation}
where $c_0$ is a constant.  Any non-trivial function $G(z)$ will generate a frictional slowing force on the beam, and continuity of the potential shape is imposed via boundary conditions $G(-a/2)=G(a/2)=0.$
The equation of motion in two dimensions is then
\begin{equation}
\left(\begin{array}{c}
\ddot{x}\\
\ddot{z}
\end{array}\right)_{-a<2z<a}=-\frac{2c_{0}}{M R^2}\left(\begin{array}{c}
x-G(z)\\
\partial_z G(z)(x-G(z))
\end{array}\right).
\end{equation}

\begin{figure}
\begin{centering}
\includegraphics[width=0.95\columnwidth]{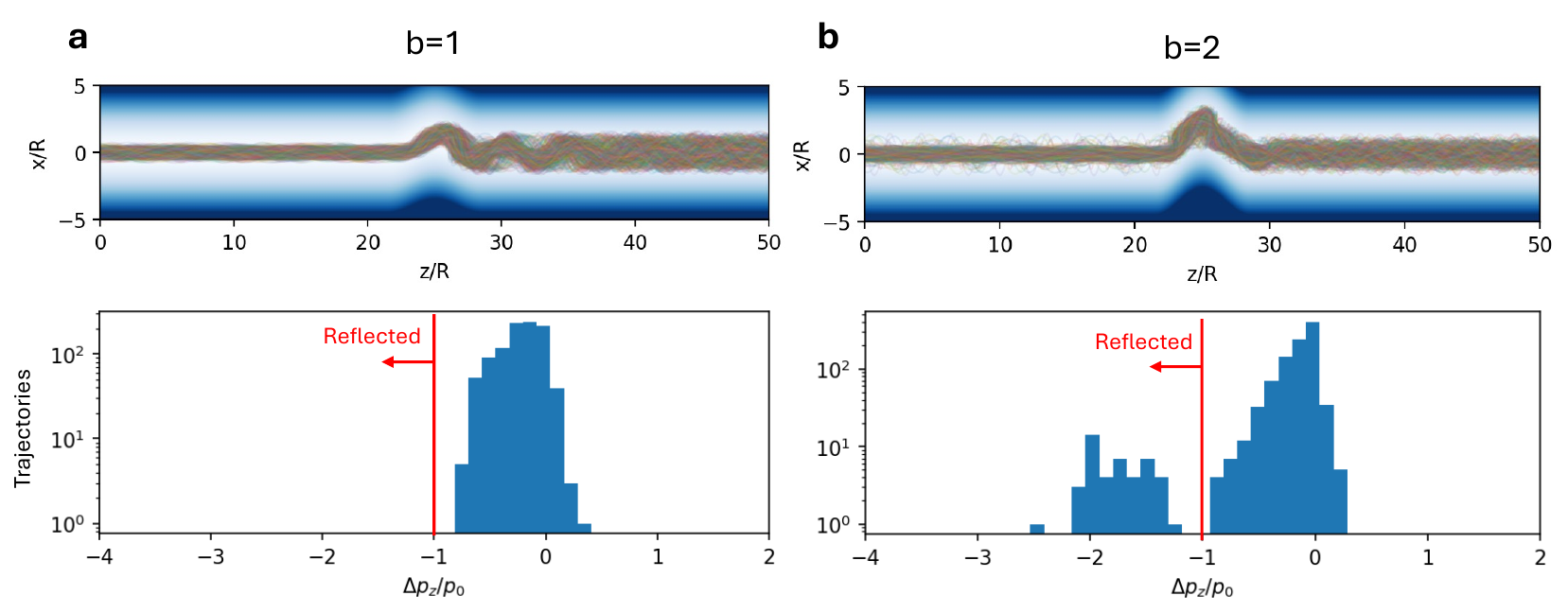}
\par\end{centering}
\caption{Demonstration of the effect of reflected particles from a perturbation
that is too large. The top panels are presented similarly to Fig.~\ref{fig:Effect-of-a}, and the bottom ones show the final $z$ momentum distribution to highlight the appearance of reflected trajectories with $-\Delta p_z>p_0$.  Figure (a) 
Shows an appropriate, small perturbation through
which all particles are transmitted. (b) Shows
a larger perturbation of the same shape, which is strong enough to
reflect a significant fraction of the particles. The eventual slowing
geometry needs to be carefully tuned to avoid these reflected particles,
as they can eject the forward-going beam from the guide. \label{fig:Demonstration-of-the}}
\end{figure}

As an example perturbation, we consider the following
 model whereby there is a single kink of controllable size dependent
on free parameter $b$.
\begin{equation}
G(z)=b\cos^{2}\left(\pi z/a\right),
\end{equation}
In this case the equations of motion become
\begin{equation}
\left(\begin{array}{c}
\ddot{x}\\
\ddot{z}
\end{array}\right)_{-a<2z<a}=-\frac{2c_0}{MR^2}\left(\begin{array}{c}
x-b\cos^{2}\left(\pi z/a\right)\\
-\frac{\pi}{a} b \sin\left(2\pi z/a\right)\left(x-b\cos^{2}\left(\pi z/a\right)\right)
\end{array}\right).\label{eq:EqOfM}
\end{equation}
The ensemble behavior for $\mathcal{M}=2.5$ in a model with $c_0/(MR^2)=10$ and $b=R$ is compared against the unscattered $b=0$ scenario in Fig.~\ref{fig:Effect-of-a}. We see
that at the location of the perturbation, the phase space distribution
is disturbed, redistributing some bulk momentum from the $\hat{z}$
direction into $\hat{x},\hat{y}$ directions. Since
each trajectory in the ensemble hits the scattering perturbation with
a different phase, each emerges with a differently perturbed transverse
momentum. This leads to expansion of the phase space distribution,
resulting in an increase in temperature of the beam. This increase
in temperature is accompanied by a reduction in the mean $p_{z}$
per particle. More dramatic perturbations also lead to reflections
of particles back in the $-z$ direction. The transverse momentum
that is developed in the guide is damped as the various particles
propagate with distinct wavelengths, eventually returning to $\left<p_{x}\right>=0$,
with reduced $\left<p_{z}\right>$ and increased internal temperature.

We can consider a slowing and cooling beamline as a sequence of slowing
perturbations and cooling segments. Rather than stringing cooling
and slowing components in series, it is conceivable that the two processes
could be arranged to occur simultaneously at all positions. An important complication 
is that immediately after the kink, the phase space distribution is
strongly pushed out of local thermal equilibrium; if the next kink
follows too soon then there will be interference effects between them, leading  to back scattering of particles upstream.  This is an inherently problematic situation, since hard collisions of the beam with those particles will lead to rapid particle losses and heating.
This effect is also evident if any individual perturbation is made
too large, as shown in the example of Fig.~\ref{fig:Demonstration-of-the}.
A geometry whereby the perturbations are separated by straight segments
of rethermalization in a linear section is one straightforward way
to ensure the beam is well conditioned when entering each perturbation,
to avoid backscattering effects. We adopt this approach
in the geometry outlined in Sec. \ref{sec:MECB-Geometry-Proposal}.

\begin{figure}
\begin{centering}
\includegraphics[width=0.95\columnwidth]{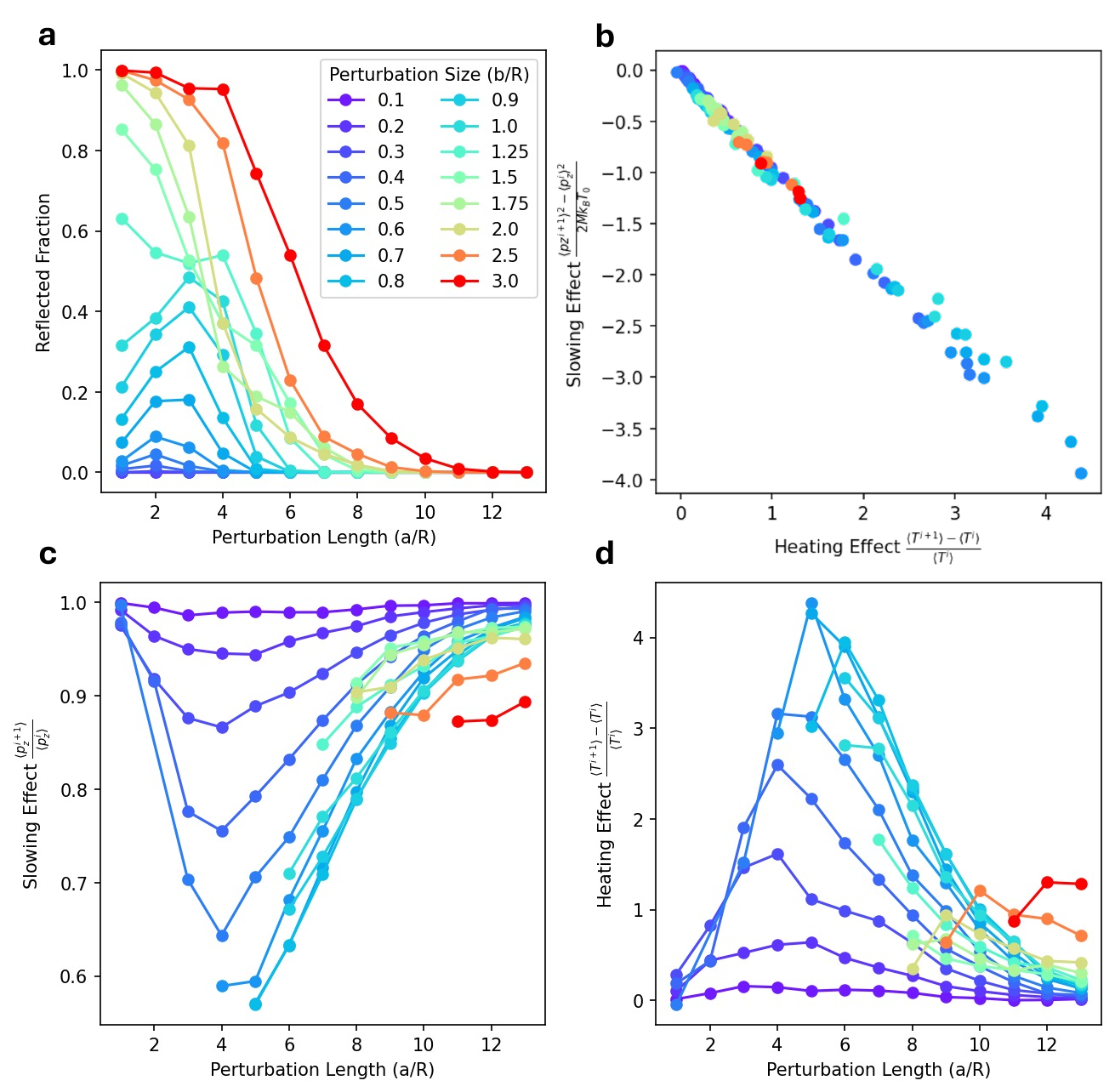}
\par\end{centering}
\caption{Parameter scan for perturbation for cooling. 
In each plot, the horizontal axis represents the perturbation length and the color denotes the perturbation size, defined in the top left caption.  The panels show: (a) reflected fraction; (b) relation between heating and slowing effects, reflecting conservation of energy; (c) slowing effect for each perturbation configuration; (d) heating effect for each perturbation configuration.  Only parameter points with less than 2\% reflected fraction are shown. \label{fig:Parameter-scan-for}}
\end{figure}

We can quantify the effect of a given guide perturbation
on beam slowing using three figures of merit: 1) the change in temperature $T$, assessed by the root-mean-square
of the momentum RMS($p$); 2) the change in bulk 
momentum $\left<p_{z}\right>$; 3) the fraction of reflected
trajectories. These can be investigated as a function of the perturbation
length $a$ and magnitude $b$ of Eq. \ref{eq:EqOfM}. Plots
showing the effects in these figures of merit in a series of calculated
geometries using 10,000 simulated trajectories drawn from a distribution
with $\mathcal{M}=2.5$ are shown in Fig.~\ref{fig:Parameter-scan-for}(a), (c), and (d), respectively. Fig.~\ref{fig:Parameter-scan-for}(b)
 shows the correlation between the slowing and cooling effects,
showing they are directly linearly related, as expected
- while scattering from the potential perturbation does not conserve
momentum, it must conserve energy, and all the energy lost from the
mean beam momentum must be absorbed into the internal motion of the
particles. That the points on Fig.~\ref{fig:Parameter-scan-for}(b)
do not fall exactly on a straight line reflects that the distribution
emerging from the kink is not a perfect Maxwell Boltzmann distribution,
as it will only assume this form after rethermalization through collisions.

Several further conclusions can be drawn from consideration of the
dependence of the slowing efficiency on perturbation parameters shown
in Fig.~\ref{fig:Parameter-scan-for}. First, as the perturbation becomes
very long, both the slowing and/or heating effects and reflection probabilities
tend to zero; in this case the perturbation simply becomes
an adiabatic curve to the guide that particles will follow without
their trajectories being significantly interrupted. The cooling effect
and reflection probabilities also tend to zero as $b\rightarrow0$,
expected since this is a limit in which the perturbation vanishes. Furthermore, for a given perturbation size $b$
there is a minimal length $a$  that does not create a significant
particle reflection effect. Any configuration with a large reflected particle
fraction is problematic for controlled cooling since those particles will
return and interact with the oncoming beam. Addition of a large number of  reflected particles also makes the longitudinal momentum distribution bimodal, 
complicating interpretation of the beam temperature and mean momentum as characteristic variables.
As such, we have opted only to plot the points in top right and the
bottom left and right panels of Fig. \ref{fig:Parameter-scan-for}
that have a small reflection probability (<2\% for these plots, though
in practice most of these points do not show any reflected trajectories). 

Based on these studies, the optimal slowing effect occurs around $a\sim5$ 
and $b\sim1$, giving a 60\% slowing and factor of 4 reheating.
These optima are found to depend slightly on the $z$ momentum of the
beam entering the perturbation, though in the range $\mathcal{M}\sim2-3$
relevant to our designs, only a small dependence is observed.  In the following calculations we will assume a 75\% slowing factor, corresponding to a factor of $\sim$3 reheating per segment. Although somewhat more aggressive slowing configurations are possible, this operating point appears to provide a suitable compromise between efficient slowing and the difficulties associated with reflected trajectories and rapid reheating of the beam.

\section{Beamline Geometries for Li and T Cooling}

We will now outline two example MECB systems, focusing on production of cold beams of atomic Li and T.  Sec.~\ref{sec:MECB-Geometry-Proposal}  presents the Li MECB test system that is currently in production for Project 8.  Sec.~\ref{sec:A-speculative-geometry} presents a speculative geometry for T cooling.

\subsection{MECB geometry for the Li pathfinder beam \label{sec:MECB-Geometry-Proposal}}

The cooling geometry proposed here anticipates a flux emerging from
an upstream element, for the case of Li a partially cooled beam
from a hot oven followed by a resonant Zeeman slower \cite{bowden2016adaptable,garwood2022hybrid,marti2010two,vanhaecke2007multistage},
that will have a mean energy significantly above the trapping strength
of the strongest available magnets. The first part of the beamline
is thus considered to be a ``catcher'', which essentially extracts the bottom part of the Maxwell Boltzmann distribution that can
be trapped. No cooling happens in this section, and the beam density
is deliberately kept low until the hot atoms have escaped by using
a large diameter, high multipolarity segment. 

\begin{figure}
\begin{centering}
\includegraphics[width=1\columnwidth]{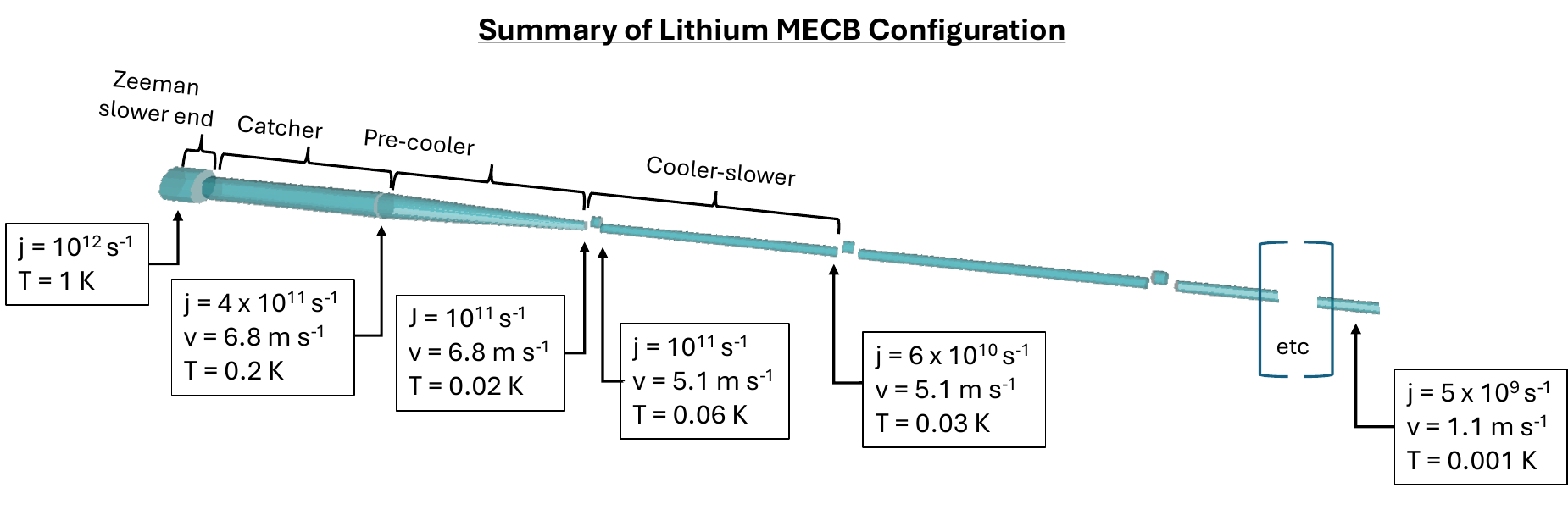}
\par\end{centering}
\caption{Schematic overview of the proposed Li MECB pathfinder geometry.  An initial catcher section captures the beam and injects it into a pre-cooler, trapping it using the strongest available permanent magnets. A sequence of six slowing and cooling segments then follows to produce a cold and slow beam.
\label{fig:Schematic-overview-of}}
\end{figure}

Once the hot atoms have left the distribution after approximately one meter
of catcher section
the beam is then brought into a condition whereby evaporative cooling is effective.
This means tuning the magnetic geometry such that the thermalization parameter
$\zeta<1$ in the catcher section becomes $\zeta\gtrsim1$ in the
evaporative cooling sections. The first stage of MECB cooling then uses the strongest available
magnets until their cooling effect is saturated and the current and temperature
stabilize. By the end of this segment the beam is well conditioned
into a moving but thermalized Maxwell Boltzmann distribution.

The remainder of the beamline after the pre-cooler is multiple meters
long and is arranged as a series of slowing and cooling units, which
evaporatively cools the beam to the required 1 mK and slows the distribution
to a low speed such that it can be injected into a trap. Avoiding zero velocity is important, as otherwise the atoms do not travel further along the beamline; as such, in designing the desired
cooling trajectories we have been careful to maintain $\mathcal{M}>\sqrt{\pi/6}$, the effusion limit.
The design shown here keeps this mass flow number $\mathcal{M}$ between $\sim1$ and $\sim2.3$,
though modifications may be made to this balance based on observed
performance in prototypes. In the Li test system, the beam is
probed to measure its current and temperature at an end station and assess
the effectiveness of the MECB design. A sketch of this beamline concept
is shown in Fig. \ref{fig:Schematic-overview-of}. 

\subsubsection{Catcher}

The goal of the catcher segment is to trap as large a fraction of the current
emerging from the Zeeman slower system as possible, by selecting
the relevant low energy part of the Maxwell-Boltzmann distribution with
the strongest possible trapping magnets. The density in this segment
is deliberately kept low, in order to avoid rethermalization while
the fastest transverse particles leave the beam. Past Zeeman slowers
\cite{stack2011ultra} have produced outgoing currents of $10^{15}$
per second at the oven and $10^{11}$ per second at the exit aperture. The Project 8 Li test source has a much larger oven aperture
than these past devices and so we expect to be able to achieve larger
currents by one to two orders of magnitude. As such, for the following
calculations we assume the incoming current to the catcher to be $j_{0}\sim10^{12}\mathrm{~s}^{-1}$,
at a pre-cooled temperature of $T_0\sim1 K$. The surviving particle fraction after a transverse evaporation cut at momentum $p_{cut}$ is then

\begin{equation}
f_{N}=\frac{\int_{0}^{p_{cut}}dp\,\left(pe^{-p^{2}/2Mk_BT}\right)}{\int_{0}^{\infty}dp\,\left(pe^{-p^{2}/2Mk_BT}\right)}=1-e^{-p_{cut}^{2}/2Mk_{B}T}=1-e^{-\eta},
\end{equation}
and the remaining energy fraction is

\begin{equation}
f^{<\eta}_{E}=\frac{\int_{0}^{p_{cut}}dp\,\left(p^{3}e^{-p^{2}/2Mk_BT_0}\right)}{\int_{0}^{\infty}dp\,\left(p^{3}e^{-p^{2}/2Mk_BT_0}\right)}=1-e^{-p_{cut}^{2}/2Mk_{B}T_0}\left(1+\frac{p_{cut}^{2}}{2Mk_{B}T_0}\right)=1-e^{-\eta}(1+\eta).
\end{equation}
Equipartition tells us the temperature scales as $T_0\propto E/N$,
which means that following this cut the remainder of the distribution
rethermalizes and will have a temperature of 

\begin{equation}
T\rightarrow\frac{f^{<\eta}_{E}}{f^{<\eta}_{N}}T_{0}=\left[1-\frac{\eta}{(e^{\eta}-1)}\right]T_{0}.
\end{equation}
If we are taking a small subset from the bottom end of the distribution,
it is appropriate to Taylor expand these expressions at small $\eta$
to obtain a particle fraction and temperature fraction
of the original following

\begin{equation}
\frac{N}{N_{0}}\rightarrow\eta,\quad\frac{T}{T_{0}}\rightarrow\frac{\eta}{2}.
\end{equation}

Strong available permanent magnets have
$B_{\rm max}=0.5$~T, implying $\eta\sim0.4$ at the top end in a simple multipole geometry, giving an effective
temperature of 0.2 K and current of $j_{\rm catcher}\sim0.4 j_{0}\sim4\times10^{11}\mathrm{~s}^{-1}$. Notably, Halbach geometries may allow for enhancement of this surface field strength, though at the cost of additional magnetic elements that would exacerbate the difficulty of removing residual gases (and eventually, in the tritium application, recombined T$_2$ molecules) from the beamline.
Since only forward-going particles will enter the catcher, the forward
velocity of this ensemble will correspond to the mean forward Maxwell Boltzmann
velocity of $v\sim 6.8~\mathrm{m~s}^{-1}$.  

\begin{figure}
\begin{centering}
\includegraphics[width=1\columnwidth]{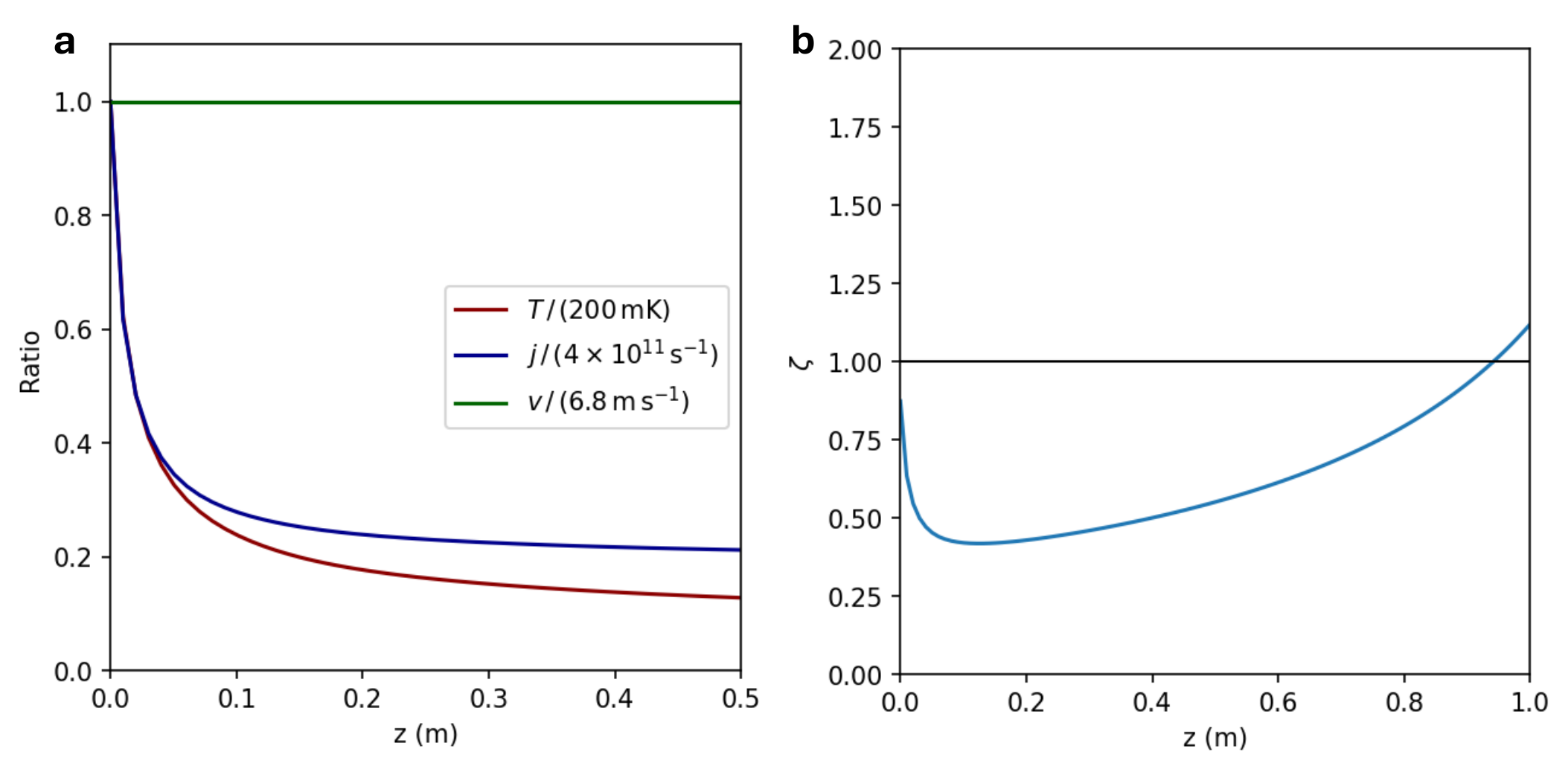}
\par\end{centering}
\caption{Initial cooling in the pre-cooler section. (a) shows the evolving temperature, current and mean velocity; and (b) shows the beam opacity parameter $\zeta$ as a function of position along the segment.  \label{fig:Initial-cooling-in}}
\end{figure}

\subsubsection{Pre-cooler}

To begin true evaporative cooling, the aforementioned current must be compressed into a thermalization
radius of around 1 cm for $\zeta\sim1$ to achieve a beam with sufficient density  for evaporative cooling. This can be accomplished via an octupole
guide of steadily reducing diameter. We have considered here an segment with
radius steadily decreasing from 3~cm to 1~cm, which provides sufficient
space to insert the appropriate bar magnets but condenses the beam
as much as possible given the available magnetic field strengths. This
also provides the first stage of evaporative cooling and conditions
the beam for the subsequent more aggressive slower-cooler section.   For the purposes of this study, we assume that the beam has become thermalized upon entering this segment, though in practice this may be partially occurring during the upstream section of the pre-cooler without significant changes to our predictions.

The calculated evaporative cooling rates in the pre-cooler are presented in Fig.~\ref{fig:Initial-cooling-in}(a). We see the current will stabilize in a $B_{max}=0.5$ T guide after
around 0.5 m with around 25\% of the incoming current and 10\% of the original
temperature, given an incoming beam with the properties predicted
above. The value of $\zeta$ in this segment can be calculated as,

\begin{equation}
\zeta=\frac{2\sqrt{2}\mu_a B_{max}\sigma j}{k_BT\pi v\Gamma\left[\frac{m+2}{m-2}\right]R}.
\end{equation}
The geometry chosen here maintains $0.5\leq\zeta\leq1.2$ all the way along
the pre-cooler, ensuring both efficient thermalization and effective
evaporative dynamics.

\subsubsection{Cooler-slower }

At the end of the pre-cooler we expect $v=6.8$~m s$^{-1}$, $j\sim 10^{11}\,\mathrm{s^{-1}}$
and $T=20\mathrm{\,mK}$, implying $\mathcal{M}\sim 2.3$. With such an incoming
current, a `slower' segment can be implemented that preserves the total
current, reduces central momentum by a factor of $\frac{3}{4}$
and increases temperature by a factor of approximately $3$.  The temperature can then be reduced using a segment of evaporative cooling at constant $\eta$ to return to the original mass flow number.  The goal of our design will thus be to achieve $\mathcal{M}\sim2.3$
at the entrance to each slowing element $i$, while evaporatively
cooling in segments with $\zeta\geq0.5$ in between. This is accomplished
in the following set of repeating steps:

\begin{figure}
\begin{centering}
\includegraphics[width=1\columnwidth]{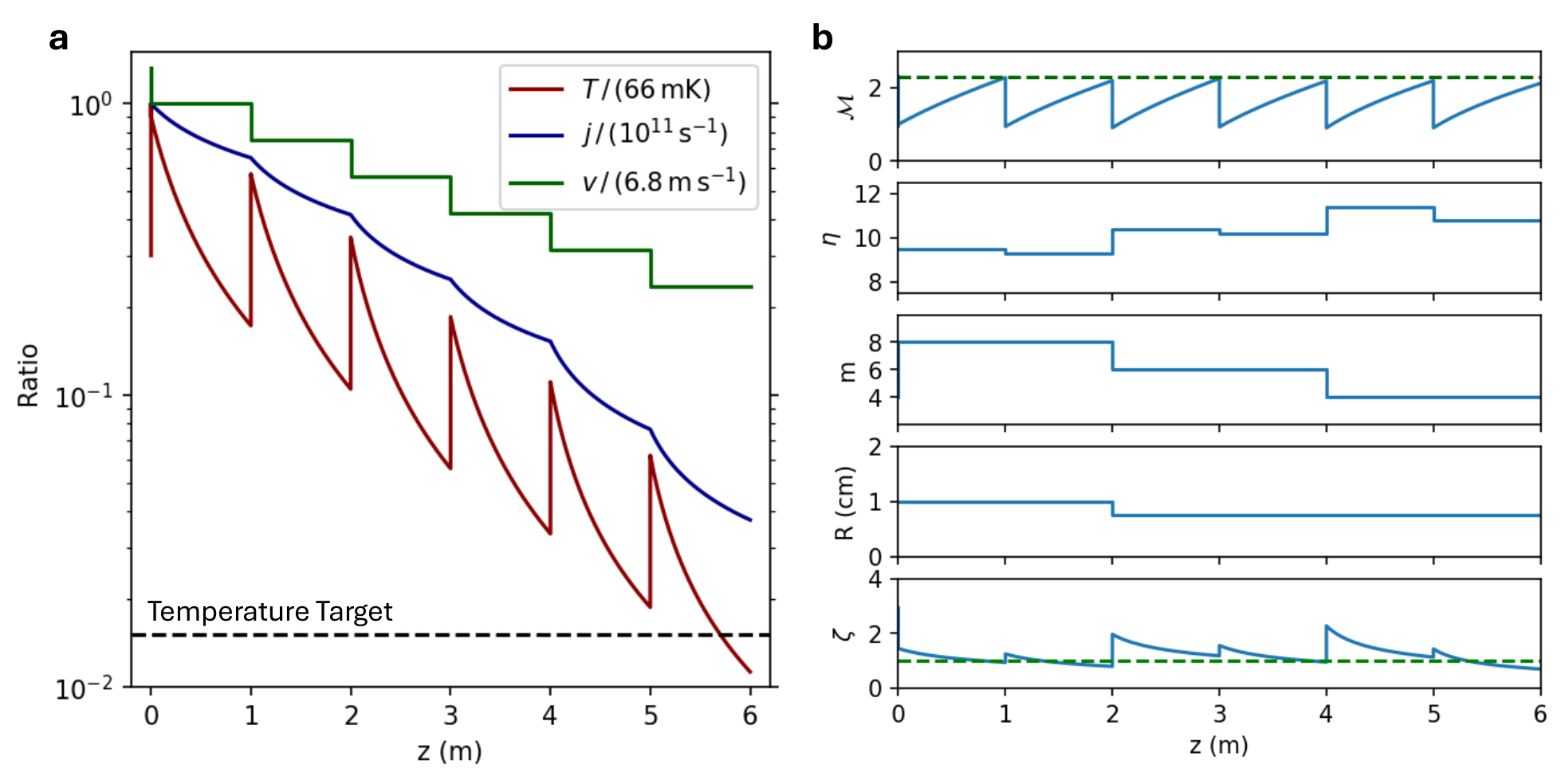}
\par\end{centering}
\caption{(a) Temperature, current and velocity evolution in proposed segmented
cooler-slower geometry for $^{6}$Li. (b) Operating parameters of this system.
$\zeta$ is kept as close to 1 as possible, while maintaining guide
dimensions of diameter at least 1.5~cm for octupoles and 1~cm for sextupoles.
These dimensions appear realistic given available magnet dimensions
with the necessary surface field strengths.   \label{fig:Cooler-with-no}}
\end{figure}

\begin{enumerate}
\item Slowing section with momentum change $p_{i+1}\rightarrow\frac{3}{4}p_{i}$ accompanied by reheating of $T_{i+1}\rightarrow3T_{i}$, 
\item Cooling section to return to $\mathcal{M}=2.3$ with temperature change $T_{i+2}\rightarrow\frac{9}{3\times 16}T_{i}$ and bulk momentum preserved as
$p_{i+2}\rightarrow p_{i+1}$
\end{enumerate}
The combined effect of these two elements is $T_{i+2}\rightarrow\frac{9}{16}T_{i}$
and $p_{i+2}\rightarrow\frac{3}{4}p_{i}$, and the sequence always
maintains at least 4 times the central momentum relative to RMS fluctuations of velocity in $z$
to keep the beam moving forward along the beamline. The choice to
slow first rather than cool first serves to minimize the total length
of the beamline, and improve thermalization in the cooling segments. 

To accomplish the goal of cooling from $\mathrm{20\,mK}$ to $1\,\mathrm{mK}$
we will need $\frac{\log[1~mK/20~mK]}{log[9/16]}\sim6$ segments. Practical considerations limit the beamline to
7~m in length, 
so we select as a design driver that each segment will be 1~m in
length. The values of $\eta$, $R$ and $m$ for each segment are
chosen to maintain as well as possible $\zeta$ at order 1, while
cooling with the required power 
in 1 m of distance.
Within the lithium beamline the flux is sufficiently low that
maintaining $\zeta\sim1$ is an important design challenge to maintain thermalization along the evaporating beam.

We predict that the proposed configuration will achieve a beam with
final temperature at 1 mK with longitudinal velocity of 1.2 ms$^{-1}$ at
5\% efficiency in the cooler-slower section. The operating parameters
of the system as well as its performance in terms of $T,$ $j$
and $v$ is shown in Fig. \ref{fig:Cooler-with-no}(a). A more efficient
beam could also be produced with longer segments and slower cooling
if desired, though the beamline length would be longer for reaching
the same target temperature. This system will allow for investigation
of the MECB cooling and slowing dynamics, in order to prepare a detailed
design for an atomic T cooling beam.

Several salient features of the MECB concept are visible in the detailed parameter trajectories of Fig.~\ref{fig:Cooler-with-no}(b). First, we see that the transition from higher multipoles to lower ones, as well as reducing the multipole diameter along the beam line, is required to keep $\zeta\sim 1$ as the beam current is reduced. This results in an accompanying decrease of cooling efficiency since higher multipoles  maintain a larger cooling exponent $\gamma$. The trade-off between these factors, as well as practical considerations such as maintaining multipole diameters at experimentally accessible scales, dictates the range of possible configurations for an effective MECB system. We have fixed the initial atom current in these calculations to values similar to those demonstrated in past experiments with Zeeman slowers. If higher incoming currents were achievable, higher multipoles would remain viable for the downstream sections, avoiding the largest losses and increasing the efficiency.  This example shows one possible set of parameters for cooling a $^{6}$Li beam, though clearly there is significant flexibility to adjust multipolarity $m$, radii $R$,  evaporation cut $\eta$, and segment lengths, all in continuous and distance-dependent ways, as well as slowing power per segment and modularity of the cooling/slowing steps.  The system is thus highly tunable with many degrees of freedom that can be traded off and optimized within the constraints imposed by considerations associated with realization of a practical geometry.

\begin{figure}
\begin{centering}
\includegraphics[width=0.6\columnwidth]{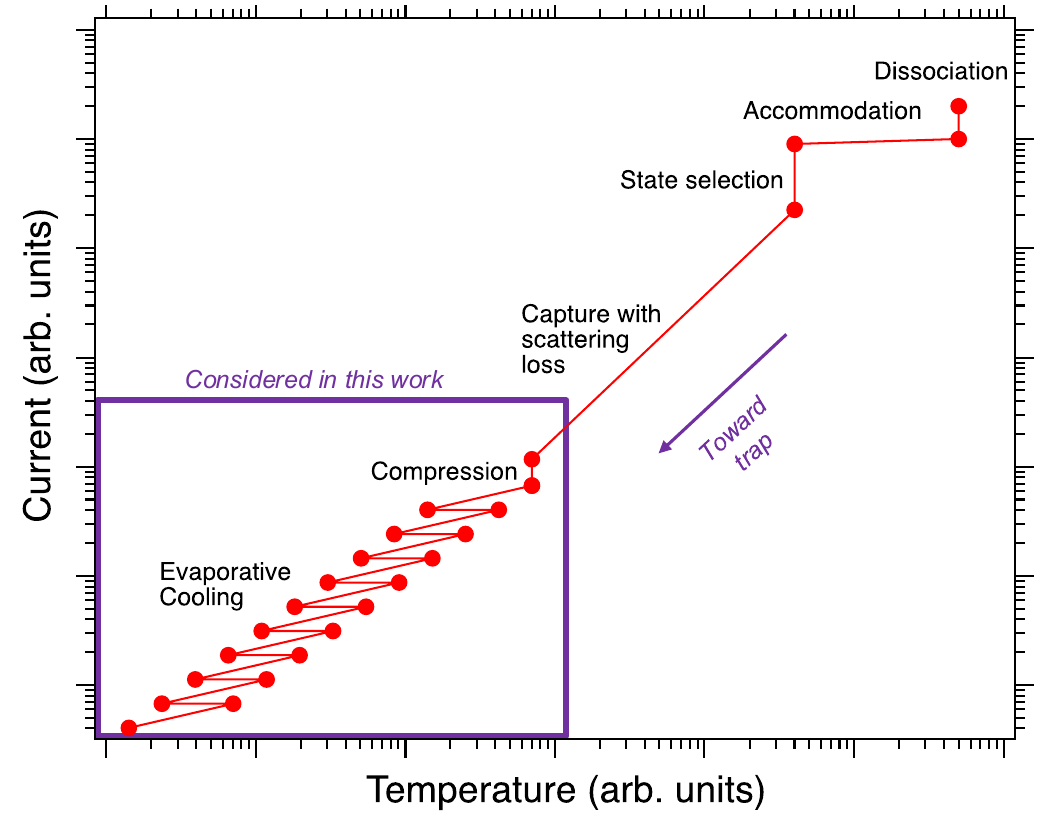}
\par\end{centering}
\caption{Schematic outline of Project 8 atomic tritium cooling scheme.  The cooling trajectory follows the red line from right (at the source) to left (at the trap). \label{fig:Tritium-system-scheme}}
\end{figure}

\subsection{Magnetic geometry for T cooling and slowing \label{sec:A-speculative-geometry}}

For the eventual T beamline, the scattering cross sections are
smaller, but the currents and velocities are higher than the lithium test system. This both affects
the dynamics of the evaporative cooling, and requires a commensurately
longer beamline. Here we use the methods of the above
section to outline a rough scheme for the T cooling system.

Projections suggest that between 10$^{14}$ and 10$^{16}$ atoms per second must be delivered to the magnetic trap at mK temperatures in order to conduct a sensitive neutrino mass search in Project 8. We thus consider here a target current of 10$^{15}$ s$^{-1}$ and target temperature of 1~mK as a representative benchmark.  The MECB line is only one of many elements that must be developed to supply cold atomic T for Project 8.  A sketch of the anticipated scheme is shown in Fig.~\ref{fig:Tritium-system-scheme}.  Molecular gas enters a dissociator that uses thermal or radio-frequency methods to crack the gas into neutral atoms at a temperature in the range 500 to 2500 K.  The atomic gas can be cooled via accommodation with a cold surface in the range 10 to 30 K.  Such temperatures are still too high for magnetic entrainment, and only the colder tail of the distribution is captured into the beamline. The losses associated with capture are substantial and carry large uncertainties. If the beam density is too small for evaporative cooling, additional losses are associated with compression.   We defer upstream tritium source considerations for future work. These upstream components in the tritium system substitute for the Li oven, Zeeman slower, catcher and pre-cooler of the Li system.

As in the lithium system, the upper end of the magnetic cooling beamline segment for T is limited by the strength of
the strongest obtainable permanent magnets.  We assume $B_{\rm max}=0.5$
T. It is notable that in a Halbach configuration, surface fields of 0.8 T can be reached~\cite{walstrom2009magneto}, though despite the apparent opportunities to reach higher trapping fields, the primary design-driving concern at the top end of the beamline is the capability to pump away recombined and warm T$_2$ molecules from the guide.  This appears to restrict the upper segments to be simple permanent quadrupole magnets with open spaces, despite the performance penalty implied in terms of trapping and cooling power.  Still higher fields could potentially  be obtained with superconducting magnets.  These are not considered in the default design on the grounds of cost and complexity, though they could be explored in future investigations.

 For present purposes we consider a current of 5$\times10^{17}$ atoms per second at an effective internal temperature of 0.2 K entering the MECB system. We also assume the bulk velocity of this distribution corresponds to effusive flow at 0.2 K, which implies a velocity of $30\mathrm{~m~s}^{-1}$. These are speculative baseline assumptions, to be met with data from test systems and more detailed calculations concerning the upstream elements.

\begin{figure}
\begin{centering}
\includegraphics[width=0.99\columnwidth]{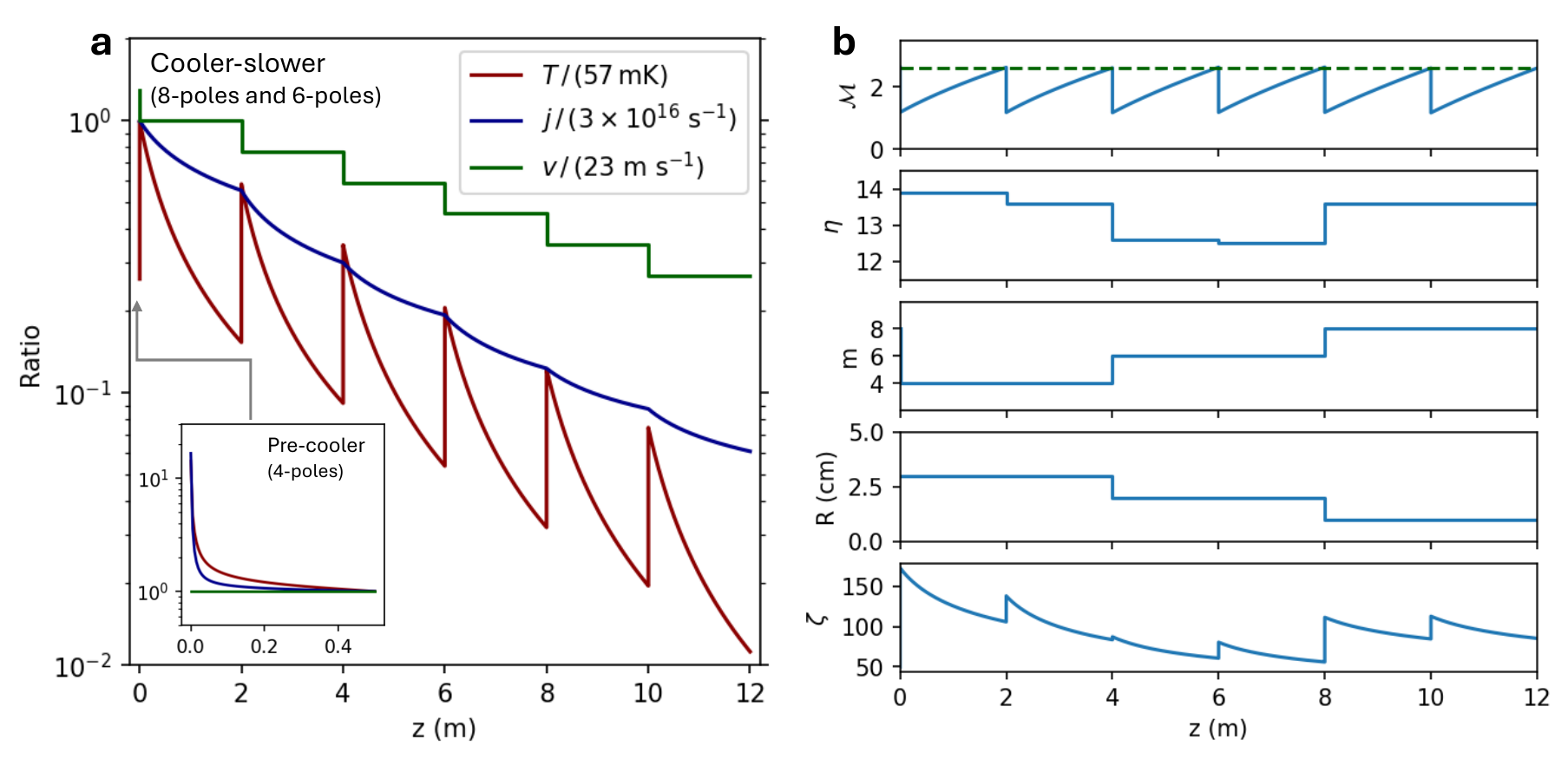}
\par\end{centering}
\caption{T system performance parameters. (a) Shows the temperature, current and mean velocity evolution along the beamline; (b) shows the evolving mass flow number $\mathcal{M}$, evaporation cut $\eta$, multipolarity $m$, multipole radius $R$ and beam opacity parameter $\zeta$.   \label{fig:Tritium-system-performance}}
\end{figure}

Both because it is important to pump out recombined T$_{2}$ gas and because maintaining $\zeta$ for thermalization is a  less pressing concern in the high current environment of the T beam than in the Li example, we use larger multipole radii in the T beam than were used there. A reversed configuration of multipolarities is able to maintain an sufficiently large value for $\zeta$.  A pre-cooler segment with quadrupole radii reducing from 6~cm to 3~cm over 0.5~m delivers 3$\times 10^{16}$~s$^{-1}$ of atoms at an effective temperature of 15~mK to the cooler-slower section. There, 3~cm radius quadrupoles transition  toward 2~cm octupoles as the atoms travel down-beam.

Over 12~m it seems to be plausible to cool the distribution to 1 mK with efficiency $10^{-3}$, using realistic parameters and assuming the T triplet scattering cross
sections outlined in Sec. \ref{subsec:Triplet-scattering-cross}.
One example geometry with its associated performance parameters is
shown in Fig.~\ref{fig:Tritium-system-performance}. In this geometry, $\zeta$ is maintained at between to 60 and 150 along the beamline, which is sufficient to thermalize the beam but not so large as to completely inhibit evaporation due to beam opacity.  The mass flow number is returned to $\mathcal{M}\sim 2.6$ following each cooling/slowing sector, using a similar scheme to that used in Sec.~\ref{sec:MECB-Geometry-Proposal}. As may be seen in Fig.~\ref{fig:Tritium-system-performance}(a) the current and temperature track each other in nearly 1-to-1 proportion.  

Majorana spin flip transitions ~\cite{majorana1932atomi} will be active along the central beam axis where the B field of the multipoles is low, leading to additional losses. These  can be suppressed by addition of a solenoidal field in addition to the multipole field.  The largest Majorana losses are expected in the quadrupoles, since these have the largest gradients in the central, low-field region.  Using methods outlined in Ref.~\cite{brink2006majorana} the expected Majorana loss rates in these segments have been calculated and are found to be less than 1\% per meter for a 20~mT applied background field.  This solenoid field strength can be relaxed significantly as the beam temperature falls and the trap field strength is reduced.

A detailed engineering design for the Project 8 MECB system, as well as a truly robust performance estimate, will require consideration of a number of other factors, including tritium handling, evacuation of warm gas from the beam, understanding of Majorana spin-flip losses in the beam, energy- and magnetic-field dependent cross sections, detailed treatment of position-dependent beam opacity, momentum dependences of scattering cross sections including partial wave contributions above s-wave, and dipole losses from d-state collisions, among others.  Nevertheless, based on these
early studies,  it appears there are realistic possibilities to use the MECB method as a central component of the Project 8 cold atomic tritium source.  

\section{Conclusions \label{sec:Conclusions}}

Dynamic evaporative cooling is a technique for achieving cold beams of magnetically trappable atoms in a laser-less configuration. The Project 8 collaboration is developing this method to provide a well-conditioned, cooled and slowed beam of atomic T to feed the Project 8 magnetogravitational trap.  The method is to be experimentally probed via an initial Li test-beam phase. 

The dynamics of a magnetic evaporative cooling beamline can be understood on the basis of thermalization and evaporation of a trapped, self-interacting atomic vapor.  A slowing component can be added via perturbations whose form is controlled by the requirement that they not reflect a substantial component of the beam, separated by cooling segments to remove the heat that is generated by the slowing process.

We have presented in this paper a suite of analytic and numerical calculations developed to estimate the performance of MECB cooling and slowing in geometries conceived for cooling of both Li and T systems.  Our projections suggest that the required currents of 10$^{14}$~s$^{-1}$-10$^{16}$~s$^{-1}$ of T at temperatures of order 1~mK may be deliverable to the Project 8 trap using the MECB approach as long as upstream elements provide a tritium current of $5\times 10^{17}$~s$^{-1}$ at 200~mK.  The ongoing atomic tritium R\&D program of Project 8 and its Li pathfinder phase will confront these estimates with data in the near future. Proof of successful MECB-based slowing and cooling will be an important step toward a direct neutrino mass measurement based on trapped atomic tritium.

\section*{Acknowledgments}

This material is based upon work supported by the following sources: the U.S. Department of Energy Office of Science, Office of Nuclear Physics, under Award No.~DE-SC0020433 to Case Western Reserve University (CWRU), under Award No.~DE-SC0011091 to the Massachusetts Institute of Technology (MIT), under Field Work Proposal Number 73006 at the Pacific Northwest National Laboratory (PNNL), a multiprogram national laboratory operated by Battelle for the U.S. Department of Energy under Contract No.~DE-AC05-76RL01830, under Early Career Award No.~DE-SC0019088 to Pennsylvania State University, under Award No.~DE-SC0024434 to the University of Texas at Arlington, under Award No.~DE-FG02-97ER41020 to the University of Washington, and under Award No.~DE-SC0012654 to Yale University; the National Science Foundation under Grant No.~PHY-2209530 to Indiana University, and under Grant No.~PHY-2110569 to MIT; the Karlsruhe Institute of Technology (KIT) Center Elementary Particle and Astroparticle Physics (KCETA); Laboratory Directed Research and Development (LDRD) 18-ERD-028 and 20-LW-056 at Lawrence Livermore National Laboratory (LLNL), prepared by LLNL under Contract DE-AC52-07NA27344, LLNL-JRNL-871979; the LDRD Program at PNNL; Yale University; and the Cluster of Excellence "Precision Physics, Fundamental Interactions, and Structure of Matter" (PRISMA+ EXC 2118/1) funded by the German Research Foundation (DFG) within the German Excellence Strategy (Project ID 39083149).

\section*{Appendix A: Numerical solution of the Boltzmann collision integral for cooling rates\label{sec:AppendixA}}

The time derivative of the phase space density required to calculate evaporative cooling rates
can be evaluated via the full collisional Boltzmann transport equation.
This equation has the full form

\begin{equation}
\frac{\partial}{\partial t}f(\vec{r},\vec{p})=-\frac{\vec{p}}{m}.\nabla_{r}f(\vec{r},\vec{p})+\nabla_{r}U.\nabla_{p}f(\vec{r},\vec{p})+{\cal I}(\vec{r},\vec{p}),
\end{equation}
where the terms represent, from left to right: time evolution of phase
space, mass flow due to inertia, force from applied potential, and
collision integral that leads to thermalization. For spherical collisions
with constant cross section, ${\cal I}$ can be written as
\begin{equation}
{\cal I}(r,\vec{p}_{4})=\frac{\sigma}{\pi m^{2}}\int d^{3}p_{3}\left[\int d^{3}p_{1}d^{3}p_{2}\right]\delta_{p}\delta_{E}\left\{ f(\vec{r},\vec{p}_{1})f(\vec{r},\vec{p}_{2})-f(\vec{r},\vec{p}_{3})f(\vec{r},\vec{p}_{4})\right\} .
\end{equation}
\begin{equation}
\delta_{p}^{3}\delta_{E}=\delta^{3}(\vec{p}_{1}+\vec{p}_{2}-\vec{p}_{3}-\vec{p}_{4})\delta\left(\frac{p_{1}^{2}}{2m}+\frac{p_{2}^{2}}{2m}-\frac{p_{3}^{2}}{2m}-\frac{p_{4}^{2}}{2m}\right).
\end{equation}
Inside the square brackets $p_{3}$ and $p_{4}$ are treated as fixed,
and we can rewrite the $p_{1}$ and $p_{2}$ integrals as
\begin{equation}
\int d^{3}p_{1}\int d^{3}p_{2}=\int d^{3}P\,d^{3}q,
\end{equation}
\begin{equation}
\vec{P}=\vec{p}_{1}+\vec{p}_{2},\quad\quad\vec{q}=\frac{\vec{p}_{1}-\vec{p}_{2}}{2}.
\end{equation}
The momentum conserving delta function $\delta_{p}^{3}$ fixes the
center of mass momentum before and after, as $\vec{p}_{1}+\vec{p}_{2}=\vec{P}=\vec{p}_{3}+\vec{p}_{4}$:
\begin{equation}
{\cal I}(\vec{r},\vec{p}_{4})=\frac{\sigma}{\pi m^{2}}\int d^{3}p_{3}\left[\int d^{3}P\delta^{3}(\vec{P}-\vec{p}_{3}-\vec{p}_{4})\right]\left[\int d^{3}q\delta_{E}\right]\left\{ f(\vec{r},\vec{p}_{1})f(\vec{r},\vec{p}_{2})-f(\vec{r},\vec{p}_{3})f(\vec{r},\vec{p}_{4})\right\} .
\end{equation}
The energy conserving delta function fixes the magnitude of $q$,
but its angle is free,
\begin{equation}
{\cal I}(\vec{r},\vec{p}_{4})=\frac{\sigma}{\pi m^{2}}\int d^{3}p_{3}\int d^{3}P\delta^{3}(\vec{P}-\vec{p}_{3}-\vec{p}_{4})\int q^{2}dq\,d\Omega_{q}\delta_{E}\left\{ f(\vec{r},\vec{p}_{1})f(\vec{r},\vec{p}_{2})-f(\vec{r},\vec{p}_{3})f(\vec{r},\vec{p}_{4})\right\} .
\end{equation}
Manipulating the energy delta function we find the $\Omega_{q}$ dependence
drops out (as expected from kinematics in the center of mass frame)
so
\begin{equation}
\delta_{E}=\delta\left(\frac{p_{3}^{2}}{2m}+\frac{p_{4}^{2}}{2m}-\frac{P^{2}}{4m}-\frac{q^{2}}{m}\right)=\frac{m}{2q}\delta\left(q-\sqrt{\left(\frac{\vec{p}_{3}-\vec{p}_{4}}{2}\right)^{2}}\right).
\end{equation}
We find that $|\vec{q}|=\frac{1}{2}|\vec{p}_{3}-\vec{p}_{4}|$ and
we have to integrate over directions. Continuing with the Boltzmann
equation,
\begin{eqnarray}
{\cal I}(\vec{r},\vec{p}_{4})&=&\frac{\sigma}{2\pi}\int d^{3}p_{3}\int d^{3}P\delta^{3}(\vec{P}-\vec{p}_{3}-\vec{p}_{4})\left[\int dq\,d\Omega_{q}\delta\left(q-\frac{1}{2}|\vec{p}_{3}-\vec{p}_{4}|\right)\right] \nonumber \\
&& \qquad \times \frac{q}{m}\left\{ f(\vec{r},\vec{p}_{1})f(\vec{r},\vec{p}_{2})-f(\vec{r},\vec{p}_{3})f(\vec{r},\vec{p}_{4})\right\} .
\end{eqnarray}
The two delta functions eliminate the freedom in $P$ and $|q|$,
leaving only the solid angle integral for $\vec{q}$,
\begin{equation}
{\cal I}(\vec{r},\vec{p}_{4})=\frac{\sigma}{2\pi}\int d^{3}p_{3}d\Omega_{q}\left[\frac{q}{m}\left\{ f(\vec{r},\vec{p}_{1})f(\vec{r},\vec{p}_{2})-f(\vec{r},\vec{p}_{3})f(\vec{r},\vec{p}_{4})\right\} \right]_{\rm constrained},
\end{equation}
where ``constrained'' means we must only evaluate at kinematically
valid points, defined by
\begin{equation}
\begin{array}{c}
\vec{P}=\vec{p}_{1}+\vec{p}_{2}=\vec{p}_{3}+\vec{p}_{4},\\
\vec{q}=\frac{1}{2}(\vec{p}_{1}-\vec{p}_{2})=\frac{1}{2}(\vec{p}_{3}-\vec{p}_{4}).
\end{array}
\end{equation}
\begin{center}
\begin{figure}
\begin{centering}
\includegraphics[width=0.9\columnwidth]{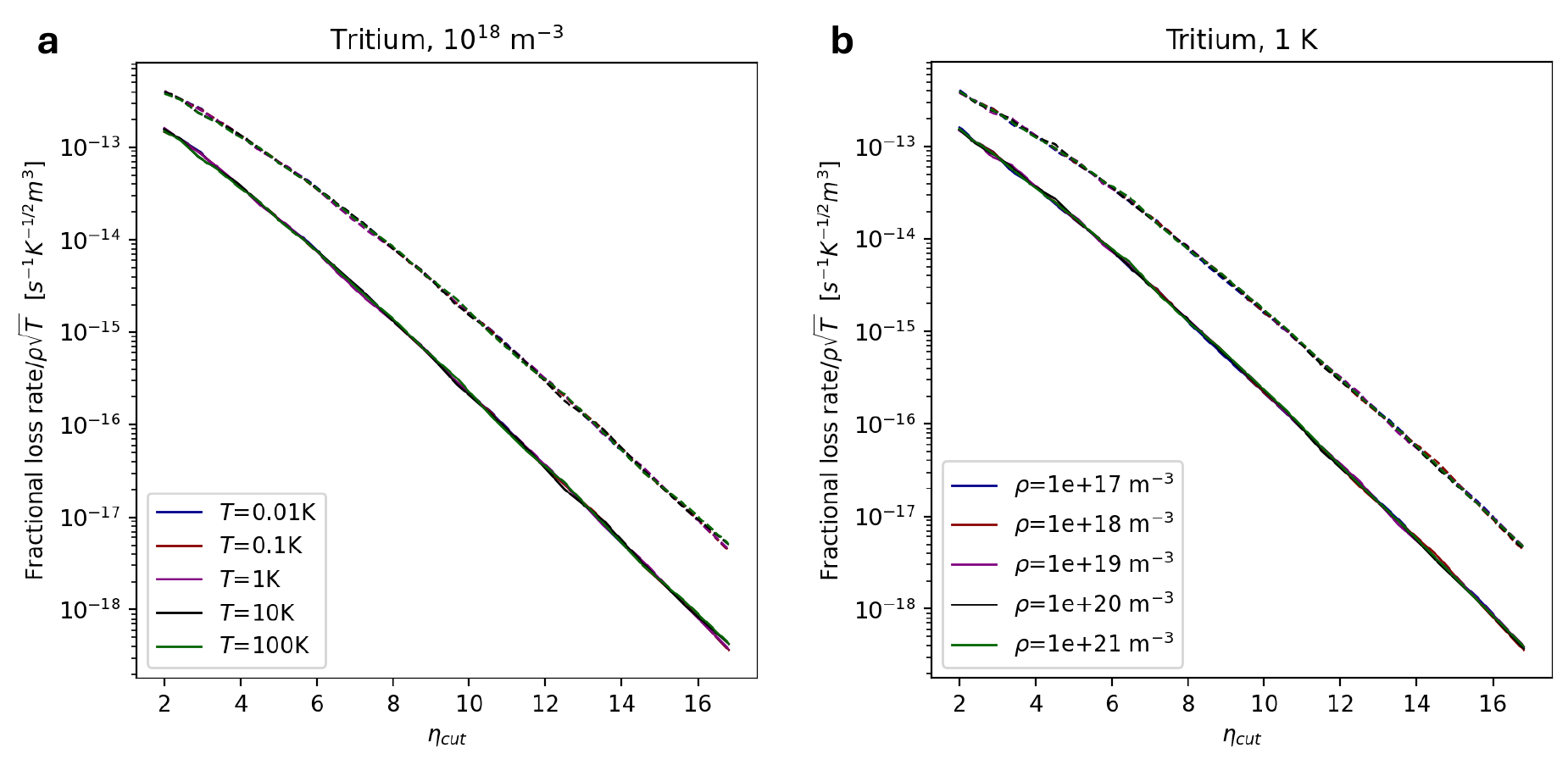}
\par\end{centering}
\caption{Scaling behavior of the $\Gamma_{E}$ (dotted) and $\Gamma_{N}$ (solid) values with
density (left) and temperature (right) demonstrating the expected
scaling behavior of Eq. \ref{eq:Scaling}\label{fig:Scaling-behaviour-of}.}
\end{figure}
\par\end{center}
We evaluate the final integrals by Monte Carlo integration over the
phase space distribution. Sampling $N$ points in $p_{3}$ and solid
angle of $q$ with $i=1...N$ from phase space volume $V$, the appropriate
discrete approximation to the integral is
\begin{equation}
\int_{V}f(\vec{x})d^{3}x=\frac{V}{N}\sum_{i}^{N}f(x_{i}).
\end{equation}
The volume of phase space runs from $p_{3,a}=-\eta_{a}...\eta_{a}$ for $a=1,2,3$.  Sampling $N$ values for $p_{3}$ and $\Omega_{q}$ from these
ranges,
\begin{equation}
{\cal I}(\vec{r},p_{4})=\frac{\sigma}{2\pi}\frac{4\pi(2\eta_{a})^{3}}{N}\sum_{i}^{N}\frac{q^{i}}{m}\left\{ f(\vec{r},p_{1}^{i})f(\vec{r},p_{2}^{i})-f(\vec{r},p_{3}^{i})f(\vec{r},p_{4})\right\} .
\end{equation}
The collision integral will also be cylindrically symmetric, so
\begin{equation}
{\cal I}(\vec{r},p_{4})\equiv{\cal I}(\vec{r},p_{4})=\frac{\sigma}{2\pi}\frac{4\pi(2\eta_{a})^{3}}{N}\sum_{i}^{N}\frac{q^{i}}{m}\left\{ f(\vec{r},p_{1}^{i})f(\vec{r},p_{2}^{i})-f(\vec{r},p_{3}^{i})f(\vec{r},p_{4})\right\}. 
\end{equation}

In principle we can re-evaluate ${\cal I}$ for any phase space distribution.
However, simple scaling arguments can be made that show that for a
distribution which is close in form to the Maxwell Boltzmann, the
collision terms have expected scaling with $\rho,\sigma,T$ of
\begin{equation}
\Gamma_{E}=\frac{\sigma}{\sigma_{0}}\frac{\rho}{\rho_{0}}\sqrt{\frac{T}{T_{0}}}\Gamma_{E}^{0}(\eta_{cut}),\quad\quad\Gamma_{N}=\frac{\sigma}{\sigma_{0}}\frac{\rho}{\rho_{0}}\sqrt{\frac{T}{T_{0}}}\Gamma_{N}^{0}(\eta_{cut}).\label{eq:Scaling}
\end{equation}
This scaling is validated using our numerical code, with results shown
in Fig.~\ref{fig:Scaling-behaviour-of}, and the agreement is excellent.
It thus suffices to numerically evaluate $\Gamma_{E}^{0}(\eta_{cut})$
and $\Gamma_{N}^{0}(\eta_{cut})$ at a representative $\sigma_{0},\rho_{0},T_{0}$
and at each $\eta_{cut}$, then map to each other value with the above
scaling laws. From this information we can explore the loss rates
of energy and particles geometrically across a multipole guide.

\newpage

\section*{Appendix B: Table of variable names\label{sec:AppendixB}}
\centering
\resizebox{!}{4.5in}{
\begin{tabular}{|c|c|}
\hline 
Symbol & Meaning\tabularnewline
\hline 
\hline 
Position & $\vec{x}=(x,y,z)$\tabularnewline
\hline 
Momentum & $\vec{p}=(p_{x},p_{y},p_{z})$\tabularnewline
\hline 
Particle mass & $M$\tabularnewline
\hline 
Boltzmann constant & $k_{B}$\tabularnewline
\hline 
Magnetic moment (atom, electron, nucleus) & $\mu_{a,e,N}$\tabularnewline
\hline
Temperature & $T$\tabularnewline
\hline 
Cooling exponent & $\gamma$\tabularnewline
\hline 
Scattering length & $\tilde{a}$\tabularnewline
\hline
Scattering cross section & $\sigma$\tabularnewline
\hline 
Scattering mean free path  & $\lambda$\tabularnewline
\hline 
Collisions per beam crossing & $\zeta$\tabularnewline
\hline 
Dimensionality of trap & $d$\tabularnewline
\hline 
Multipolarity & $m$\tabularnewline
\hline 
Potential power law exponent & $\nu$\tabularnewline
\hline 
Mass flow number  & ${\cal M}$\tabularnewline
\hline 
Phase space function & $f(\vec{x},\vec{p})$\tabularnewline
\hline 
Density & $\rho(\vec{x})$\tabularnewline
\hline 
Total particle number  & $N$\tabularnewline
\hline 
Energy & $E$\tabularnewline
\hline 
Scaled energy ($E/k_{B}T$) & $\epsilon$\tabularnewline
\hline 
Density of energy states  & $g(E)$\tabularnewline
\hline 
Chemical potential & $\tilde{\mu}$\tabularnewline
\hline
Beam longitudinal kinetic energy  & $K$\tabularnewline
\hline 
Evaporative heat loss & $\dot{Q}_{\rm ev}$\tabularnewline
\hline 
Logarithmic cooling rates (energy, number) & $\Gamma_{E}, \Gamma_{N}$\tabularnewline
\hline 
Fraction of (energy, number), that are (above, below) cut $\eta$ & $f_{E/N}^{>\eta/<\eta}$\tabularnewline
\hline 
Multipole guide barrier height (units of $k_{B}T$) & $\eta$\tabularnewline
\hline 
Evaporation cut (units of $k_{B}T$) at a specific position & $\eta_{cut}$\tabularnewline
\hline 
Excess energy loss per evaporated particle (units of $k_{B}T$) & $\eta'$\tabularnewline
\hline 
Momentum at evaporation cut  & $p_{\rm cut}$\tabularnewline
\hline 
Magnetic field & $B$\tabularnewline
\hline 
Surface $B$ field of a single permanent magnet & $B_{0}$\tabularnewline
\hline 
Maximum $B$ field at surface within multipole& $B_{max}$\tabularnewline
\hline 
Multipole segmentation correction to $B$ field & $\alpha(m)$\tabularnewline
\hline 
Radial position & $r$\tabularnewline
\hline 
Beam bulk velocity  & $v$\tabularnewline
\hline 
Flux (particles per second through a surface) & $j$\tabularnewline
\hline 
Radius of multipole & $R$\tabularnewline
\hline 
Radius in which majority of particles are contained & $R_{\rho}$\tabularnewline
\hline 
Trapping potential  & $V$\tabularnewline
\hline 
Slowing perturbation parameters & $a$, $b$, $c_0$\tabularnewline
\hline 
\end{tabular}
}

\newpage 

\bibliographystyle{ieeetr}
\bibliography{biblio}

\end{document}